\documentclass[12pt]{article}
\usepackage{amsmath}
\usepackage{amssymb}
\usepackage{amsfonts}
\usepackage{graphics}
\usepackage[dvips]{color}
\def\beq{\begin{equation}}
\def\eeq{\end{equation}}
\def\beqarray{\begin{eqnarray}}
\def\eeqarray{\end{eqnarray}}
\def\bsig{\mbox{\boldmath$\sigma$}}

\def\btheta{\mbox{\boldmath$\theta$}}

\def\brho{\mbox{\boldmath$\rho$}}

\def\bLam{\mbox{\boldmath$\Lambda$}}
\def\bnabla{\mbox{\boldmath$\nabla$}}
\def\sumint{\sum\mspace{-25mu}\int}

\begin{document}
{\bf
\begin{center}
Poincar\'e Invariant Quantum Theory
\end{center}
\begin{center}
J. Lab Lectures
\end{center}
\begin{center}
W. N. Polyzou
\end{center} 
\begin{center}
Department of Physics and Astronomy,
\end{center}
\begin{center}
University of Iowa
\end{center}
\begin{center}
Iowa City, IA 52242
\end{center}
}
\bigskip

\vfill\eject

\section{Background:}  

Local quantum field theory is the theory of choice for modeling
reactions involving energy and distance scales where relativity and
quantum theory both have to be considered.  Quantum Chromodynamics
(QCD) is generally accepted to be the quantum field theory that
governs the strong interactions.  Given this, it is natural to ask 
why one might consider an alternative theoretical approach.  

The short answer is that we do not really know how to solve QCD, or
any non-trivial local four-dimensional field theory.  From a practical
point of view it is not known how to compute ab-initio error bounds to
any field-theoretic calculations.  No one has been tempted to dismiss
QCD just because a ``QCD prediction'' does not agree with experiment.
From a mathematical point of view, even the existence of QCD is a
Millennium problem\cite{clay}.  This is why non-relativistic potential
models are still used for realistic computations.

One of the difficulties with all local field theories is due to the
requirement that the theory be local.  This condition means that
``experiments'' done in arbitrarily small space-like separated
spacetime regions should be independent.  This sounds like a sensible
requirement, however formulating this condition requires a theory with
an infinite number of degrees of freedom that is defined on all energy
scales.  This leads to many of the infinities that make the theory so
difficult to define.  This can be fixed by introducing cutoffs, but
this leads to violations of locality on some scale, and it is known
\cite{St65} that violations on one scale lead to violations on all scales.
Establishing that the cutoffs can be removed in a controlled manner
leaving a non-trivial theory with the expected properties is the
difficult problem. 

The point is that the field theory does a lot more than is needed to
realistically describe a class of reactions at some finite energy
scale.  The key properties of a field theory that should survive at
finite energy scales are the quantum mechanical interpretation,
Poincar\'e invariance, cluster properties, and a spectral condition.
In addition, the dynamics should dominated by a finite
number of degrees of freedom. 

While it is a central goal of nuclear physics to resolve all of these
issues with QCD, one still wants to develop a quantitative
understanding of classes of phenomena that dominate the physics at a
given energy scale.  Effective field theory does this for low-energy
reactions, but it is not generally applicable in the few GeV region.
At higher energies quasipotential approaches, such as the Gross
equation\cite{fg}, provide a phenomenology that is motivated by
quantum field theory.  Poincar\'e invariant quantum theory is a
framework for constructing quantum mechanical models of systems of a
finite number of degrees for freedom consistent with Poincar\'e
invariance, cluster properties (for fixed numbers of particles), 
and a spectral condition \cite{schroer},

It is suitable for describing reactions in the few GeV region, which
are still dominated by a relatively small number of degrees of
freedom.  There is an inverse scattering theorem \cite{baum}, so there
are interactions that can describe any reaction.  Cluster properties
provide relations between reactions involving different numbers of
degrees of freedom.  Poincar\'e invariant quantum mechanics is a
minimal extension of the standard potential theory that is
successfully being used to obtain a quantitative understanding of
low-energy physics to energies in the few GeV range.  While there is
no direct connection to QCD, one expects that QCD could be used to
provide insight into both the choice of relevant degrees of freedom
and an understanding of the operator structures that appear in the
interactions and current operators.

\section {Quantum theory}

Quantum mechanics is a linear theory.  The mathematical setting for a
quantum theory is a complete complex linear vector space, or Hilbert
space.  I represent vectors in the Hilbert space by ``kets''
\beq
\vert \psi \rangle .
\eeq
The Hilbert space inner product in Dirac's Bra-Ket notation
has the familiar form:
\beq
\langle \psi \vert \phi \rangle .
\eeq
All of the predictions of a quantum theory are expressed in terms of 
this inner product.   In what follows I assume that the Hilbert space 
vectors are normalized to unity:
\beq
\langle \psi \vert \psi \rangle =1 .
\eeq
The quantities that are measured in a quantum theory are
probabilities, expectation values, and ensemble averages.
Probabilities are given directly in terms of the Hilbert space inner
product by
\beq
P_{\phi,\psi} = \vert \langle \phi \vert \psi \rangle \vert^2 .
\eeq
This represents the probability of measuring a system
to be in the state represented by the vector $\vert \phi \rangle$ if it
was initially prepared in the state represented by the vector $\vert \psi
\rangle$.  

A related quantum observable is the expectation value of a Hermitian
operator $A$ in the state $\vert \psi \rangle$.  Hermitian operators
have a complete set of eigenvectors, $\vert n \rangle$, with real
eigenvalues $a_n$.  The {\color{red}expectation value} of $A$ in the
state $\vert \psi \rangle$,
\beq
\langle \psi \vert A \vert \psi \rangle =
\sum_{n}  \vert \langle \psi \vert n \rangle \vert^2  a_n
=  \sum_{n} P_{n,\psi} a_n ,
\eeq
is the weighted average of the eigenvalues of $A$ by the probabilities of
measuring the system to be in the $n$-th eigenstate of $A$ if it is
initially prepared in the state $\vert \psi \rangle$.

To measure probabilities or expectation values it is necessary to
perform a statistically significant number of measurements on
identically prepared initial states.  In most experiments it is
impossible to ensure the initial states are identically prepared.  For
example, a polarized beam of particles is never 100\% polarized.  Instead,
it consists of a statistical distribution of polarizations.  This
situation is treated by using an ensemble average.  In this case I
assume that the initial state is in a statistical ensemble of quantum
states $\vert \psi_m \rangle$ distributed with classical probability
$P_m$.  The expectation value of the Hermitian operator $A$ in this
ensemble is defined by
\[
\langle A \rangle = \sum_m P_m \langle \psi_m  \vert A \vert \psi_m  \rangle 
\]  
This expectation value can be expressed by the trace of the product of 
$A$ multiplied by a density 
matrix $\rho$:
\[
<A>  := \mbox{Tr} (\rho A)  \qquad \rho = \sum_m  
\vert \psi_m \rangle  P_m \langle \psi_m \vert
\qquad \sum_m P_m =1 
\]
where here the different  $\vert \psi_m \rangle$ do not 
have to be orthogonal.

An important observation is that Hilbert space scalar products and
eigenvalues of Hermitian operators are unitarily invariant. Since
quantum probabilities, expectation values, and ensemble averages are
constructed out of Hilbert space scalar products and eigenvalues of
Hermitian operators, all quantum observables are unitarily invariant.  

Abstract observables are useful in quantum field theories, but the
observables that are relevant to scattering experiments are normally
associated with isolated particles.  A complete
measurement of the state of an asymptotically free particle involves a
determination of the particle's linear momentum, up to some finite
experimental resolution in a particular coordinate system, and the
projection of its spin along some axis in the same (or a different)
coordinate system.  A {\color{red}complete experiment}
would involve a measurement of the state (momentum
and spin projection) of each of the asymptotic particles produced in
a scattering reaction.

These measurements normally involve using a combination of
conservation laws and observations of the trajectories of charged
particles in classical electromagnetic fields. 

The identification of a complete measurement provides a means to pass
from an abstract formulation of the Hilbert space of quantum theory to
an explicit representation of a model Hilbert space.  Any complete
measurement of a scattering process will give probabilities of
measuring the projection of the particle's spin along a given axis and
its linear momentum to be in a finite volume $V$ of momentum space.  Thus
vectors in a single-particle Hilbert space are square integrable 
functions of the linear
momentum and magnetic quantum number $\langle \mathbf{p}, \mu 
\vert \psi \rangle $, where
\beq
\int_V \vert \langle \mathbf{p}, \mu \vert \psi \rangle \vert^2 d^3 p 
\eeq
represents the probability that a particle in state 
$\vert \psi \rangle$ will be measured to have linear momentum in the 
momentum volume $V$ and magnetic quantum number $\mu$.
The normalization condition
\beq
\sum_{\mu} \int \vert \langle \mathbf{p}, \mu \vert 
\psi \rangle  \vert^2 d^3 p =1
\eeq
means that the probability of finding 
the particle any of its allowed states is 1.

Thus, the Hilbert for a single particle can be chosen as the
space of square integrable functions in the particle's momentum and
spin projection.  N-particle Hilbert spaces can be taken as N-fold
tensor products of single particle spaces.  Spaces describing states
with variable particle number can be taken as orthogonal direct sums
of N-particle spaces.

For example, a suitable Hilbert space to describe nucleon-nucleon 
scattering at an energy sufficient to produce no more than 
one pion can be taken to be
\beq
{\cal H} = ({\cal H}_n \otimes {\cal H}_n)
\oplus ({\cal H}_n \otimes {\cal H}_n \otimes {\cal H}_{\pi} ) 
\eeq
Wave functions have the form
\beq
\langle \cdot \vert \psi \rangle =
\left (
\begin{array}{l} 
\langle \mathbf{p}_1, \mu_1, \mathbf{p}_2, \mu_2 \vert
\psi_{nn} \rangle \\
\langle \mathbf{p}_1, \mu_1, \mathbf{p}_2, \mu_2 ,\mathbf{p}_{\pi} 
\vert
\psi_{nn\pi} \rangle
\end{array} 
\right ) 
\eeq 
with normalization 
\beq 
1= \langle \psi \vert \psi \rangle = \langle
\psi_{nn} \vert \psi_{nn} \rangle + \langle \psi_{nn\pi} \vert
\psi_{nn\pi} \rangle .  
\eeq 
The quantity 
\beq
\langle
\psi_{nn} \vert \psi_{nn} \rangle
\eeq
is the probability that the state $\vert \psi \rangle$ will 
be measured to have two nucleons and no pions and
\beq
\langle
\psi_{nn\pi} \vert \psi_{nn\pi} \rangle
\eeq
is the probability that the state $\vert \psi \rangle$ will 
be measured to have two-nucleons an one pion.  

Thus, for any bounded energy range the observable experimental
reaction products determine a representation of the Hilbert space with
sufficient structure to describe all accessible experimental
observables.  In some applications, for example with models involving
confined degrees of freedom, it is possible and useful to use
different degrees of freedom.  In any representation the experimental
degrees of freedom (physical particles) must eventually appear in the
formulation of the scattering asymptotic conditions that are used to
define scattering probabilities.

\section {Special Relativity}

A fundamental assumption of special relativity is the existence of
inertial coordinates systems.  Inertial coordinate systems have the
property that equivalent experiments done in different inertial
coordinate systems lead to identical results.

Experimentally, the Michelson-Morley experiment established that
different inertial coordinate systems are related by transformations
that preserve the proper distance (time) 
\beq 
\vert \mathbf{x} - \mathbf{y} \vert^2 -
c^2 (t_x-t_y)^2 =
\vert \mathbf{x}' - \mathbf{y}' \vert^2 -
c^2 (t'_x-t'_y)^2 . 
\label{c.1}
\eeq  
between the space-time coordinates of ``events'', which are labeled by
their space and time coordinates $(t,\mathbf{x})$ and
$(t',\mathbf{x}')$ in different inertial coordinate systems.

The Poincar\'e group is the group of transformations that preserves 
the quadratic form (\ref{c.1}).  I use 4-vectors, $x^{\mu}$, to label 
events 
\beq
x \to x^{\mu} = (x^0,x^1,x^2,x^2)= (ct,x^1,x^2,x^2) 
\label{c.2}
\eeq
and use the Minkowski metric
\beq
\eta_{\mu\nu} = \eta^{\mu \mu} =
\left ( 
\begin{array}{cccc}
-1 &0 &0 &0 \\
0 &1 &0 &0 \\
0 &0 &1 &0 \\
0 &0 &0 &1
\end{array}
\right ).
\label{c.3}
\eeq
The most general point transformation, 
$x' = f (x)$, 
satisfying (\ref{c.2}) is called
a Poincar\'e transformation which has the general form 
\beq 
x^{\mu}
\to x^{\prime \mu} = \sum_{\nu=0}^3 \Lambda^{\mu}{}_{\nu} x^{\nu} +
a^{\mu}
\label{b.4}
\eeq
where $a^{\mu}$ is constant and $\Lambda^{\mu}{}_{\nu}$ is a constant
matrix, called a Lorentz transformation, satisfying
\beq
\eta^{\mu \nu} = \sum_{\alpha \beta}
\Lambda^{\mu}{}_{\alpha} \Lambda^{\nu}{}_{\beta} \eta^{\alpha \beta} .
\label{b.5}
\eeq
In what follows I use Einstein's summation convention, which assumes
that repeated lower case Greek letters are summed from 0 to 3. 

The full Poincar\'e group is generated by space translations, time
translations, rotations, and rotationless Lorentz boosts which depend
continuously on a parameter, as well as the discrete transformations
of space reflection, time reversal, and four dimensional reflections.

It is experimentally observed that the discrete Poincar\'e
transformations are not symmetries of the weak interaction.  In what
follows the symmetry group of special relativity will be taken as the
subgroup of the Poincar\'e group where the Lorentz transformations can
be continuously deformed to the identity.  I use the term Poincar\'e
transformation to refer to this subgroup of the full Poincar\'e group.

{\it Classically} relativistic invariance is interpreted to mean that
dynamical equations remain unchanged under changes of inertial
coordinate system.  This is because the solution of the dynamical
equations is observable.  This leads to the notion that the equations
of a relativistically invariant theory should be ``covariant''.  This
is no longer necessary in a quantum theory.  

In a quantum theory the results of experimental measurements are
quantum probabilities, expectation values, and ensemble averages.
These quantities are not solutions of dynamical equations.  Special
relativity requires that these quantum observables remain invariant 
under change of inertial coordinate system.

The invariance of quantum observables with respect to changes in
inertial coordinate system means that the group of Poincar\'e
transformations continuously connected to the identity is a symmetry
of the quantum theory.  {\color{red} In 1939 Wigner \cite{Wi39} showed
that this requirement is equivalent to the existence of a unitary
representation of the Poincar\'e group on the quantum mechanical
Hilbert space.  Below are some observations related to Wigner's
theorem.} Poincar\'e invariant quantum mechanics is simply a quantum
theory with a unitary representation of the Poincar\'e group.

\begin{itemize}

\item [1.]  Wigner's results apply to any quantum theory including 
quantum field theory.  Wigner's theorem motivated all serious attempts to 
axiomatize quantum field theory \cite{St65}\cite{Haag}\cite{Os}.   

\item [2.]  As stated above, discrete Poincar\'e transformations
are not considered part of the Poincar\'e group when discussing
special relativity.

\item [3.] Antiunitary transformations do not appear because any
continuous Poincar\'e transformation can be written at the square of
another Poincar\'e transformation.

\item [4.] Wigner's unitary representations of the Poincar\'e group
are ray representations - Later Bargmann \cite{Ba54} showed 
that they could be replaced by single-valued representations 
of the covering group, which replaces Lorentz transformations 
by the group $SL(2,\mathbb{C})$.

\item [5.] Wigner's theorem says nothing about microscopic causality;
  however the Poincar\'e invariance is not compatible with the
  existence of a sensible position operator for particles
\cite{Ne49} , which would
  be needed test locality in a theory of particles.  Implementation of
  a test of microscopic causality requires additional degrees of
  freedom normally associated with fields.  The difficulty in defining
  a suitable position operator for a particle has nothing to do with
  antiparticles.

\item [6.] The basic building blocks of unitary representations of the
  Poincar\'e group are the irreducible representations, which were
  also classified by Wigner in \cite{Wi39}.  Vectors in the positive mass
  positive energy irreducible representations have exactly the same
  quantum numbers as a particle of mass $m$ and spin $s$.  This will
  be discussed in section 7.
  
\end{itemize}   

\section {Parameterization of the Poincar\'e group}

A general Poincar\'e group element $(\Lambda ,a)$ is labeled by a Lorentz
transformation $\Lambda$ and a space-time translation four vector $a$.
The group product is
\beq
(\Lambda_2 ,a_2 )(\Lambda_1 ,a_1 )= (\Lambda_2\Lambda_1 ,\Lambda_2 a_1 +a_2 ) 
\label{d.1}
\eeq
where $\Lambda_2\Lambda_1$ means $\Lambda_2^{\mu}{}_{\alpha} 
\Lambda_1^{\alpha}{}_{\nu}$ 
and  $\Lambda_2 a_1$ means $\Lambda_2^{\mu}{}_{\alpha} a_1^{\alpha} $.
The identity is
\beq
I = (I,0)
\label{d.2}
\eeq
and inverse is
\beq
(\Lambda ,a )^{-1} = (\Lambda^{-1},-\Lambda^{-1} a). 
\label{d.3}
\eeq 

The Poincar\'e group is a 10 parameter group.  There are 10
independent one-parameter subgroups associated with rotations about
the $\hat{\mathbf{x}}$, $\hat{\mathbf{y}}$ and $\hat{\mathbf{z}}$
axes, rotationless Lorentz transformations in the $\hat{\mathbf{x}}$,
$\hat{\mathbf{y}}$ and $\hat{\mathbf{z}}$ direction, translations in
the $\hat{\mathbf{x}}$, $\hat{\mathbf{y}}$ and $\hat{\mathbf{z}}$
directions and time translations.  These elementary transformation can
be used to generate any Poincar\'e transformation.

The rotations can be parameterized by an axis and angle of rotation, $\btheta$:
\beq
\Lambda \to R(\btheta) = e^{i\mathbf{L}\cdot \btheta}
\label{d.4}
\eeq
where
\beq
\mathbf{L} =
i \left ( 
\begin{array}{cccc} 
0 & 0 & 0 & 0 \\
0 & 0 & 0 & 0 \\
0 & 0 & 0 & -1 \\
0 & 0 & 1 & 0 
\end{array}
\right ),
i \left ( 
\begin{array}{cccc} 
0 & 0 & 0 & 0 \\
0 & 0 & 0 & 1 \\
0 & 0 & 0 & 0 \\
0 & -1 & 0 & 0 
\end{array}
\right ),
i\left ( 
\begin{array}{cccc} 
0 & 0 & 0 & 0 \\
0 & 0 & -1 & 0 \\
0 & 1 & 0 & 0 \\
0 & 0 & 0 & 0 
\end{array}
\right )
\label{d.5}
\eeq
while the rotationless Lorentz boost that transforms a particle of 
mass $m$ at rest to linear momentum $\mathbf{p}$ is  
\beq
\Lambda \to \mathbf{B}(\mathbf{p}/m)  =
\left ( 
\begin{array}{cc} 
h/m & \mathbf{p}/m \\
\mathbf{p}/m & I + {\mathbf{p} \otimes \mathbf{p} \over m(m+h)}
\end{array}
\right )
\label{d.6}
\eeq
where $h = \omega (\mathbf{p}) = \sqrt{\mathbf{p}^2 + m^2}$. 
This Lorentz transformation can be expressed in terms of the rapidity
$\brho$:
\beq
\hat{\brho} = \hat{\mathbf{p}} \qquad
{\vert \mathbf{p} \vert \over m} = \sinh (\vert \rho \vert) 
\qquad
{\vert \omega(\mathbf{p})  \vert \over m} = \cosh (\vert \rho \vert) 
\label{d.7}
\eeq
which plays a similar role as the angles in the rotation group.
When expressed in terms of angles or rapidity, 
$\mathbf{B}(\mathbf{p}/m) \to \mathbf{B}(\brho)$,
rotations about a specific axis or rotationless 
Lorentz transformations in given direction become one 
parameter groups 
\beq
R(\theta_1 \hat{\mathbf{n}}) R(\theta_2 \hat{\mathbf{n}}) = 
R((\theta_1+ \theta_2)  \hat{\mathbf{n}})  
\qquad
B(\rho_1 \hat{\mathbf{n}}) B(\rho_2 \hat{\mathbf{n}}) = 
B((\rho_1+ \rho_2)  \hat{\mathbf{n}})  
\label{d.8}
\eeq
Lorentz transformations can also be represented by $2\times 2$ complex
matrices with determinant $=1$\cite{St65}\cite {Wi60}.  This
representation is useful for computations and well as for establishing
general properties of the Lorentz group.  To motivate this
representation note that any four vector can be expressed as a $2
\times 2$ Hermitian matrix as follows:
\beq
X = x^{\mu} \sigma_{\mu} = 
\left ( 
\begin{array}{cc} 
x^0 +x^3 & x^1-i x^2 \\
x^1+i x^2 & x^0 -x^3 
\end{array} 
\right ) 
\qquad
x^{\mu} = {1 \over 2} \mbox{Tr} (X \sigma_{\mu}) 
\label{d.9}
\eeq
where $\sigma_0=I$ and $\sigma_i$ are the three Pauli matrices.
The connection with the Lorentz group follows because 
\beq
\mbox{det} (X) = - \eta_{\mu \nu} x^{\mu} x^{\nu} = -x^2 . 
\eeq
Real Lorentz transformations correspond to linear transformations that
preserve this determinant and the Hermiticity of $X$.  Up to
irrelevant constant multiplicative factors, the most general
transformation with these properties is
\beq
X' = AXA^{\dagger}  \qquad \mbox{det} (A)=1 .
\label{d.10}
\eeq
It follows from (\ref{d.9}) that 
\beq
\Lambda^{\mu}{}_{\nu} = {1 \over 2} \mbox{Tr} (\sigma_{\mu}
A \sigma_{\mu} A^{\dagger} ).
\label{d.11}
\eeq
The most general $A$ with this property has the form
\beq
A = \pm e^{{1\over 2} (\brho + i \btheta ) \cdot \bsig}
\label{d.12}
\eeq
where when $\btheta$ is zero the transformation is a rotationless
boost with rapidity vector $\brho$ and when $\brho$ is zero this
transformation is a rotation with angle of rotation $\btheta$.  Both
$A$ and $-A$ correspond the same Lorentz transformations. In general
there is a 2 to 1 correspondence with all Lorentz transformations
continuously connected to the identity.  In this representation boosts
are positive (negative) Hermitian matrices with determinant 1 and
rotations are unitary matrices with determinant 1.

\section{Unitary representations}

Unitary representations of the Poincar\'e group are unitary operators 
$U(\Lambda ,a)$ satisfying the group representation property
\beq
U(\Lambda_2 ,a_2 )U(\Lambda_1 ,a_1 )= U(\Lambda_2\Lambda_1 ,\Lambda_2 a_1 +a_2 )
\label{e.1}
\eeq
\beq
U(I,0) = I
\label{e.2}
\eeq
\beq
U[(\Lambda ,a )^{-1}] = U(\Lambda^{-1},-\Lambda^{-1} a)= U^{\dagger}
(\Lambda ,a ). 
\label{e.3}
\eeq 

The Poincar\'e group has 10 independent one parameter subgroups
(\ref{d.8}) labeled by angle of rotation (3), rapidity of a Lorentz boost (3) , 
spatial displacement (3), and temporal displacement (1).
The generators of these transformations are the Hermitian operators
\beq
\mathbf{J}\cdot \hat{\mathbf{x}}  = -i {d \over d \theta} U 
(R(\theta \hat{\mathbf{x}},0)_{\vert_{\theta = 0}}
\label{e.4}
\eeq
\beq
\mathbf{K}\cdot \hat{\mathbf{x}}  = -i {d \over d \rho} U 
(B(\rho \hat{\mathbf{x}},0)_{\vert_{\rho = 0}}
\label{e.5}
\eeq
\beq
\mathbf{P}\cdot \hat{\mathbf{x}}  = -i {d \over d a} U 
(I, a \hat{x})_{\vert_{a = 0}} 
\label{e.6}
\eeq
\beq
H  = i {d \over d t} U 
(I, (t,\mathbf{0}) _{\vert_{t = 0}}
\label{e.7}
\eeq
The generators $\{ H, \mathbf{P},\mathbf{J},\mathbf{K}\}$ have the 
familiar interpretations; $\mathbf{J}$ is the angular momentum operator,
$\mathbf{K}$ is the generator of rotationless Lorentz transformations,
$\mathbf{P}$ is the linear momentum operator, and 
$H$ is the Hamiltonian.

The group property (\ref{d.1}) implies that the generators
can be grouped into operators that 
transform
as tensors with respect to Lorentz transformations
\beq
P^{\mu} = (H, \mathbf{P} )
\label{e.8}
\eeq
\beq
J^{\mu \nu} =
\left (
\begin{array}{cccc}
0 & -K_x &-K_y & -K_z \\
K_x & 0  & J_z &  -J_y \\
K_y & -J_z  & 0 &  J_x \\
K_z &  J_y  & -J_x & 0 \\
\end{array}
\right ) 
\label{e.9}
\eeq
which have the transformation properties
\beq
U^{\dagger} (\Lambda , a) P^{\mu} U(\Lambda , a) =
\Lambda^{\mu}{}_{\nu} P^{\nu}
\label{e.10}
\eeq
\beq
U^{\dagger} (\Lambda , a) J^{\mu \nu} U(\Lambda , a) =
\Lambda^{\mu}{}_{\alpha}
\Lambda^{\nu}{}_{\beta} (J^{\alpha \beta} - a^{\alpha}P^{\beta}
+ a^{\beta}P^{\alpha} ) .
\label{e.11}
\eeq
The group representation property can be used to show that
these operators also satisfy the commutations relations
\beq
[P^{i} , P^{j}]= 0 
\qquad
[P^{i} , H]= 0 
\label{e.12}
\eeq
\beq
[J^{i} , J^j]= i \epsilon_{ijk} J^k 
\qquad 
[J^{i} , P^j]= i \epsilon_{ijk} P^k 
\label{e.13}
\eeq
\beq
[J^{i} , K^j]= i \epsilon_{ijk} K^k 
\qquad
[J^{i} , H]= 0 
\label{e.14}
\eeq
\beq
[K^{i} , K^j]= - i \epsilon_{ijk} J^k 
\qquad
[K^{i} , H]= i P^i 
\qquad
[K^{i} , P^j]= i \delta_{ij} H  .
\label{e.15}
\eeq
The spin is related to the Pauli-Lubanski vector \cite{Lu42}
\beq
W^{\mu} = {1 \over 2} \epsilon^{\mu}{}_{\nu \alpha \beta}P^{\nu} J^{\alpha \beta} 
\label{e.16}
\eeq
which satisfies
\beq
[P^{\mu} , W^{\nu}] = 0 
\qquad
[W^{\mu} , W^{\mu}] = i \epsilon^{\mu \nu}{}_{\alpha \beta} W^{\alpha} 
P^{\beta} 
\label{e.17}
\eeq
The Poincar\'e group has two polynomial invariants
\beq
M^2 = H^2 - \mathbf{P}\cdot \mathbf{P} \qquad \mbox{and} \qquad 
W^2  
\label{e.18}
\eeq
When $M^2 \not=0 $ the invariant $W^2$ can be replaced by the spin:
\beq
j^2 := W^2/M^2 . 
\label{e.19}
\eeq

\section {Irreducible representations}

Irreducible representations of the Poincar\'e group are important
in Poincar\'e invariant quantum theory because they are the elementary
building blocks of any unitary
representation of the Poincar\'e group.

Irreducible representation of the Poincar\'e group can be classified
by the eigenvalues of the invariant operators $M^2$ and $W^2$ as well
as the sign of $P^0$.  Wigner \cite{Wi39} identified six classes of
irreducible representations associated with the joint spectrum of
$M^2$ and $P^0$.  For each class there is a representative vector that
is invariant under a subgroup of the Lorentz group, called the little
group for that representation.  These are listed in Table 1.

\begin{table}
\begin{center}
\begin{tabular}{|llll|}
\hline
$M^2 >0$ & $P^0>0$ & $p_0 = (M,0,0,0)$ & $SU(2)$ \\

$M^2 >0$ & $P^0<0$ & $p_0 = (-M,0,0,0)$ & $SU(2)$ \\
    
$M^2 =0$ & $P^0>0 $ & $p_0 = (1,0,0,1)$ & $ E(2)$ \\ 

$M^2 =0$ & $P^0<0$ & $ p_0 = (-1,0,0,1)$ & $ E(2)$ \\ 

$M^2 =0$ & $P^0=0 $ & $ p_0 = (0,0,0,0)$ & $ SL(2,\mathbf{C})$ \\

$M^2 <0$ & & $ p_0 = (0,0,0,\mu)$ & $  SU(1,1) $ \\
\hline
\end{tabular}
\caption{} 
\end{center}
\end{table}

Here $E(2)$ is the covering group for the two-dimensional Euclidean 
group.  Irreducible representations of the Poincar\'e group are constructed 
by starting with irreducible representations of the little group that leaves
the canonical vector $p_0$ invariant, followed by a parameterized set of
Lorentz transformation that change the value of $p_0$  

 The most interesting representations for physics are the
representations with $M^2 >0 \qquad P^0>0$ (massive particles) and the
representations $M^2 =0 \qquad P^0>0$ (massless particles).  In these
lectures I only consider the case of massive particles.

Relativistic invariance of an isolated particle implies the existence of 
a one-body unitary representation of the Poincar\'e group.
For particles the eigenvalues of the operators 
\beq
M \qquad j^2 = W^2/M^2
\label{f.7}
\eeq
are the particle's mass $m$ and spin $j^2=j(j+1)$. 

A basis for a one-particle representation can be constructed as the
set of simultaneous eigenstates of a maximal set of commuting
Hermitian functions of the single-particle Poincar\'e generators $H$,
$\mathbf{P}$ $\mathbf{J} $, and $\mathbf{K}$.

One set of operators satisfying these conditions is:
\beq
M^2, W^2, \mathbf{P}, \hat{\mathbf{z}} \cdot \hat{\mathbf{W}} 
\label{f.8}
\eeq
It is useful to replace $W^{\mu}$ by the spin variables
\beq
j^2 = W^2/M^2 \qquad   
\label{f.9}
\eeq
and 
\beq
(0, \mathbf{j}_c) :=
{1 \over m} B (-\mathbf{p}/m)^{\mu}{}_{\nu} W^{\nu}   
\label{f.10}
\eeq
where $B (-\mathbf{p}/m)^{\mu}_{\nu}=
B^{-1}  (\mathbf{p}/m)^{\mu}_{\nu}
$ is a $4\times 4$ matrix 
of operators obtained by replacing the parameter $\mathbf{p}$ by the 
momentum operator in (\ref{d.6}):
\beq
(0, \mathbf{j}_c) := {1 \over m}
B (-\mathbf{p}/m)^{\mu}{}_{\nu} W^{\nu}   =
\label{f.11}
\eeq
\beq
(0,   {\mathbf{W} \over m} +
\mathbf{p}{\mathbf{p} \cdot \mathbf{W} + W^0 (m+h) \over m^2 (m+h)})  . 
\label{f.12}
\eeq
It follows from (\ref{e.17}) that 
\beq
[j_c^k,j_c^l]=i \epsilon^{kln} j_c^n  \qquad 
\mathbf{j}_c \cdot \mathbf{j}_c= j^2 \qquad [\mathbf{j},\mathbf{p}]=0
\label{f.13}
\eeq
A suitable set of commuting observables for a single particle is 
the mass, spin, linear momentum and $\hat{\mathbf{z}}$ component of 
the canonical spin.  

The $SU(2)$ commutation relations imply  the spin can only have
integral or half-integral eigenvalues $j$.  The spectrum of the linear
momentum operator is $\mathbb{R}^2$ because it linear momentum can be
boosted to any frame.

Thus, the Hilbert space for a particle of mass $m$ spin $j$ is 
the space of square integrable
functions of the linear momentum and spin, 
\beq
\langle \mathbf{p},\mu \vert \psi \rangle 
\qquad 1 = \int d\mathbf{p} \sum_{\mu=-j}^j 
\vert  \langle \mathbf{p},\mu \vert \psi \rangle \vert^2 =1
\label{f.14}
\eeq

The action of $U(\Lambda ,a)$ on these states is expressed in terms of 
\[
\langle \mathbf{p},\mu \vert  U(\Lambda ,a) \vert \psi \rangle =
\int \sum_{\mu'=-j}^j \langle \mathbf{p},\mu  \vert U(\Lambda ,a) \vert 
 \mathbf{p}',\mu' \rangle d\mathbf{p}' 
\langle  \mathbf{p}',\mu'
\vert \psi \rangle =
\]
\beq
\int \sum_{\mu'=-j}^j D^{jm}_{\mu , \mathbf{p} ; \mu' , \mathbf{p}'} 
(\Lambda ,a)  d\mathbf{p}' 
\langle  \mathbf{p}',\mu'
\vert \psi \rangle 
\label{f.15}
\eeq
where
\beq
D^{mj}_{\mu , \mathbf{p} ; \mu' , \mathbf{p}'} 
(\Lambda ,a):=
\langle \mathbf{p},\mu  \vert U(\Lambda ,a) \vert 
\mathbf{p}',\mu' \rangle
\label{f.16}
\eeq

It is a consequence of the definitions that
the matrices $D^{mj}_{\mu , \mathbf{p} ; \mu'' , \mathbf{p}''} 
(\Lambda ,a)$  are unitary representation of the Poincar\'e group.
\[
\int \sum_{\mu''=-j}^j d\mathbf{p}''
D^{mj}_{\mu , \mathbf{p} ; \mu'' , \mathbf{p}''} 
(\Lambda_2 ,a_2) d\mathbf{p}''
D^{mj}_{\mu'' , \mathbf{p}'' ; \mu' , \mathbf{p}'} 
(\Lambda_1 ,a_1) =
\]
\beq
D^{mj}_{\mu , \mathbf{p} ; \mu' , \mathbf{p}'} 
(\Lambda_2 \Lambda_1  ,\Lambda_2 a_1+ a_2)
\label{f.17}
\eeq
These representations are irreducible.  The matrices were derived for
a single particle of mass $m$ and spin $j$, but all positive mass
positive energy irreducible representations of the Poincar\'e group
have this form in the basis (\ref{f.14}).

In the next section I show how to construct the matrices 
\beq 
D^{mj}_{\mu , \mathbf{p} ; \mu' , \mathbf{p}'} 
(\Lambda  , a)
\label{f.18}
\eeq
for any 
$m>0$ and $j$. 
 
\section{\bf Factorization theorem}

The action of $U(\Lambda ,a)$ on the one-particle Hilbert space is
determined by the matrix $D^{mj}_{\mu , \mathbf{p} ; \mu' ,
\mathbf{p}'} (\Lambda , a)$.  To compute $D^{mj}_{\mu , \mathbf{p} ;
\mu' , \mathbf{p}'} (\Lambda , a)$ I use the following factorization
theorem:

\noindent {\bf Factorization Theorem:} Let $(\Lambda ,a)$ be any
Poincar\'e transformation and $p$ be any time-like four momenta, $p:=
(\omega_m (\mathbf{p}), \mathbf{p})$.  Then $(\Lambda ,a)$ can be
expressed as the product of (1) a rotationless Lorentz transformation
to a frame where $\mathbf{p}$ is zero, (2) a rotation, (3) a
translation of the system at rest and (4) a rotationless Lorentz
transformation to the frame with the transformed momentum, $\bLam p$:
\[
(\Lambda ,a) = 
\]
\[
(B(\bLam p/m),0) (I,B^{-1} (\bLam p/m) a) \times
\]
\beq 
( B^{-1} (\bLam p/m) \Lambda B( \mathbf{p}/m),0)  (B^{-1}(\mathbf{p}/m),0)
\label{g.1}
\eeq

The proof of this theorem follows by evaluating the above
expression using the group multiplication property.  

The factorization theorem implies
\[
U(\Lambda ,a) \vert \mathbf{p}',\mu \rangle = 
\]
\[
U[B(\bLam p'/m),0] U[I,B^{-1} (\bLam p'/m)a)] \times
\]
\beq
U[ B^{-1} (\bLam p'/m)
\Lambda B(\mathbf{p}'/m)],0] U[B^{-1}(\mathbf{p}'/m),0] \vert \mathbf{p}',\mu \rangle 
\label{g.2}
\eeq
where I use a ``prime'' to indicate an eigenvalue.

The factorization theorem reduces the computation of 
$D^{mj}_{\mu , \mathbf{p} ; \mu' , \mathbf{p}'} 
(\Lambda  , a)$ to the 
following four steps:
\begin{itemize}
\item [1.] {\bf Inverse boost to a rest state:}
{\color{red}
\beq
U[B^{-1}(\mathbf{p}/m),0] \vert \mathbf{p},\mu \rangle =
\vert  \mathbf{0},\mu \rangle \sqrt{ {m \over \omega (\mathbf{ p} )}} 
\label{g.3}
\eeq
}
\item [2.] {\bf Rotation of a rest state:}
{\color{red}
\[
U[ B^{-1} (\bLam p'/m)
\Lambda B(\mathbf{p}'/m)],0] 
\vert \mathbf{0},\mu \rangle =
\]
\beq
\sum_{\nu=-j}^j 
\vert \mathbf{0},\nu \rangle 
D^j_{\nu \mu} 
[ B^{-1} (\bLam p/m)
\Lambda B(\mathbf{p}/m) ] 
\label{g.4}
\eeq
}
\item [] where 
\beq
R_w (\Lambda ,\mathbf{p}/m) :=  B^{-1} (\bLam p/m)
\Lambda B(\mathbf{p}/m)
\eeq
\item [] is a Wigner rotation and 
\beq
D^j_{\nu \mu} [R]  = 
\langle j, \nu \vert
U(R,I) 
\vert j, \mu \rangle  
\label{g.6}
\eeq
\item [] is the standard spin $j$ irreducible representation of 
the rotation group.
\item [3.] {\bf Translation of a rest state:}
{\color{red}
\beq
U[I,B^{-1} (\bLam p/m)a)
\vert \mathbf{0},\nu \rangle  =
e^{i a \cdot \Lambda p }
\vert \mathbf{0},\nu \rangle 
\label{g.7}
\eeq
}
\item [4.] {\bf Rotationless boost of a rest state:} 
\beq
U[B (\mathbf{p}'/m),0] \vert \mathbf{0},\mu \rangle =
\vert  \mathbf{p}',\mu \rangle \vert {\omega (\mathbf{p}') \over m} ) 
\vert^{1/2} 
\label{g.8}
\eeq
\end{itemize} 

Combining these four elementary unitary transformations gives
{\color{red}
\beq
U(\Lambda ,a ) \vert (m,j) \mathbf{p},\mu \rangle
\label{g.10}
\eeq
\beq
\sum_{\nu=-j}^j 
\vert \bLam p,\nu \rangle
e^{i \Lambda p \cdot a } 
D^j_{\nu \mu} 
[ B^{-1} (\bLam p/m)
\Lambda B(\mathbf{p}/m) ] 
\vert {\omega (\bLam p ) \over \omega (\mathbf{ p} )} \vert^{1/2} 
\label{g.11}
\eeq
}
Comparing (\ref{f.15}) and ( \ref{g.11}) gives
{\color{red}
\[
D^{mj}_{\mu' , \mathbf{p}' ;
\mu , \mathbf{p}} (\Lambda , a) =
\]
\beq
\delta (\mathbf{p}'- \bLam p) 
e^{i p' \cdot a } 
D^j_{\nu \mu} 
[ B^{-1} (\bLam p/m)
\Lambda B(\mathbf{p}/m)  ] 
\vert {\omega (\bLam p ) \over \omega (\mathbf{ p} )} \vert^{1/2} 
\label{g.12}
\eeq
}
This is the mass $m$ spin $j$ irreducible representation of the 
Poincar\'e group in the $\vert \mathbf{p},\mu \rangle$ basis.
All positive-mass positive-energy irreducible representations have
this explicit form in the momentum canonical spin basis.

A second important observation is that this four step process can be
used to construct irreducible representation from any rest state that
transforms irreducibly with respect to rotations.  This will be used
to construct Clebsch-Gordan coefficients in the next section and
dynamics in the following section.

In the remainder of this section I discuss the proof of the 
elementary relations  (\ref{g.3}), (\ref{g.4}), (\ref{g.7})
and (\ref{g.8}).

The transformation properties of the linear momentum in 
equations (\ref{g.3}) and (\ref{g.8}) is a consequence of the 
transformation properties of the four momentum operator.

Invariance of the spin in (\ref{g.3}) and (\ref{g.8}) follows from 
the definition of the spin in terms of the generators and
the transformation properties of the generators:
\[
U(\Lambda ,0) \mathbf{j} U^{\dagger} (\Lambda ,0) =
\]
\[
U(\Lambda ,0) {1 \over m} B^{-1} (\mathbf{p}/m) W  U^{\dagger}(\Lambda ,0) =
\]
\[
{1 \over m} B^{-1} (\bLam^{-1} {p}/m) \Lambda^{-1} W  =
\]
\[
{1 \over m} B^{-1} (\bLam^{-1} {p}/m) \Lambda^{-1} 
B (\mathbf{p}/m) B^{-1} (\mathbf{p}/m) W =
\]
\beq
B^{-1} (\bLam^{-1} {p}/m) \Lambda^{-1} 
B (\mathbf{p}/m) \mathbf{j} 
\label{g.13}
\eeq
The operator multiplying $\mathbf{j}$ becomes a Wigner rotation when
it is applied to a momentum eigenstate.  For $\Lambda= B^{-1}
(\mathbf{p}/m)$ this Wigner rotation becomes the identity, which
implies that $\mu$ is unchanged in (\ref{g.3}) and (\ref{g.8}).

The square root factors that appear in 
(\ref{g.3}) and (\ref{g.8}) are needed to ensure that 
$U [B^{-1} (\mathbf{p}'/m),0]$ and $U [B(\mathbf{p}'/m),0]$
are unitary for states with a delta function normalization:
\beq
\langle \bLam p \vert \bLam p' \rangle  = 
\langle \mathbf{p} \vert \mathbf{p}' \rangle 
\delta (\bLam{p} - \bLam{p}') =
\delta (\mathbf{p} - \mathbf{p}') 
\vert {\partial \mathbf{p} \over \partial \bLam p
} \vert =
\langle \mathbf{p} \vert \mathbf{p}' \rangle 
\vert {\partial \mathbf{p} \over \partial \bLam p 
} \vert
\eeq
which leads to the identification 
\beq
\vert \bLam p' \rangle =
\vert \mathbf{p}' \rangle 
\vert {\partial \mathbf{p} \over \partial \bLam p
} \vert^{1/2} 
\eeq
The Jacobian 
\beq
\vert {\partial \mathbf{p} \over \partial \bLam p }\vert 
= 
{\omega (\mathbf{p}) \over \omega (\bLam p )}   
\eeq
can be read off of 
\beq
\int \delta (p^2 + m^2) d^4p  =
\int {d\mathbf{p} \over \omega (\mathbf{p} )} = 
\int {d\bLam p \over \omega (\bLam p )}   
\label{g.9}
\eeq

Equation (\ref{g.4}) follows because the transformation is a rotation
in an irreducible basis for the rotation group, while (\ref{g.7}) 
follows because the basis state is an eigenstate of the four momentum. 

\section {Clebsch-Gordan Coefficients}

The one-particle representations of the Poincar\'e group constructed
in the previous section are positive-mass positive-energy irreducible
representations of the Poincar\'e group.  In general any positive-mass
positive-energy irreducible representation has the form (\ref{g.12}).

In this section I show how to decompose a product of two irreducible
representations of the Poincar\'e group into mutually orthogonal
irreducible subspaces.  The final result looks very much like the
non-relativistic decomposition into center of mass and relative
momentum variables.  These non-interacting irreducible representations
are used in the next section to construct dynamical irreducible
representations.

The tool for performing this construction is the Clebsch-Gordan
coefficients for the Poincar\'e group.  I construct these coefficients
in this section.

Consider of tensor product of two irreducible basis vectors:
\beq 
\vert \mathbf{p}_1 , \mu_1 \rangle \otimes 
\vert \mathbf{p}_2 , \mu_2 \rangle  
\label{h.1}
\eeq

Define the kinematic variables  
\beq
P^{\mu} = \mathbf{p}_1 + \mathbf{p}_2
\label{h.2}
\eeq
\beq
M_0^2 = - \eta_{\mu \nu} P^{\mu} P^{\nu}   
\label{h.3}
\eeq
where all of the single particle momenta are on their mass shells - 
i.e. $\mathbf{p}_i^0 = \sqrt{m_i^2 + \mathbf{p}_i^2}$

I define
\beq 
k^{\mu}_i = B^{-1}(\mathbf{P}/M_0)^{\mu}{}_{\nu}p_i^{\nu} 
\label{h.4}
\eeq
and
\beq
\mathbf{k}:= \mathbf{k}_1= -\mathbf{k}_2  
\label{h.5}
\eeq

Apply 
\beq
U_0 [B^{-1}(\mathbf{P}/M_0)]:=
U_1 [B^{-1}(\mathbf{P}/M_0)] \otimes 
U_2 [B^{-1}(\mathbf{P}/M_0)]
\label{h.6}
\eeq
to the basis vector (\ref{h.1}) to get 

\[
U_0 [B^{-1}(\mathbf{P}/M_0)] 
\vert \mathbf{p}_1 , \mu_1 \rangle \otimes 
\vert \mathbf{p}_2 , \mu_2 \rangle  =
\]
\[
\vert \mathbf{k} , \mu_1' \rangle 
\sqrt{\omega_{m_1}(\mathbf{k}) \over 
\omega_{m_1}(\mathbf{p}_1)} 
D^{j_1}_{\mu_1' \mu_1} [B^{-1}(\mathbf{k}/m_1)
B^{-1}(\mathbf{P}/M_0)B^{-1}(\mathbf{p}_1/m_1)] 
\otimes
\]
\beq 
\vert -\mathbf{k} , \mu_2' \rangle 
\sqrt{\omega_{m_2}(\mathbf{k}) \over 
\omega_{m_2}(\mathbf{p}_2)} 
D^{j_2}_{\mu_2' \mu_2} [B^{-1}(-\mathbf{k}/m_2)
B^{-1}(\mathbf{P}/M_0)B^{-1}(\mathbf{p}_2/m_2)] 
\label{h.7}
\eeq
This defines a rest state of the two-particle system.  
If
\beq
\vert \mathbf{k} , \mu_1' \rangle \otimes
\vert -\mathbf{k} , \mu_2' \rangle 
\label{h.8}
\eeq
can be decomposed into irreducible 
representations with respect to rotations then the factorization theorem
can be used to construct two-particle irreducible representations.

In order to understand transformation properties
$\mathbf{k}, \mathbf{j}_1$, and  $\mathbf{j}_2$
under rotations $R$ note
\beq 
k' := B^{-1} (R \mathbf{P} /M_0) R p_1 =
B^{-1} (R \mathbf{P} /M_0) R B (\mathbf{P} /M_0) k 
\eeq
\label{h.9}
The transformation 
\beq
B^{-1} (R \mathbf{P} /M_0) R B (\mathbf{P} /M_0) 
\label{h.10}
\eeq
is a Wigner rotation of the rotation $R$.  The rotationless
boosts have the property that  
\beq
B^{-1} (R \mathbf{P} /M_0) R B (\mathbf{P} /M_0) = R
\label{h.11}
\eeq
I will prove this at the end of this section using the $2\time 2$ 
matrix representations.   Using (\ref{h.11}) in (\ref{h.9}) gives
\beq  
k' = R k 
\label{h.12}
\eeq
The spins have a similar transformation property:
\[
\mathbf{j}_1' = 
\]
\[
{1 \over m_1} B^{-1}(R \mathbf{p}_1/m_1 )
R W =   B^{-1}(R \mathbf{p}_1/m_1 )
R B(\mathbf{p}_1/m_1 ) {1 \over m_1} B^{-1}(\mathbf{p}_1/m_1 ) W =
\]
\beq
B^{-1}(R \mathbf{p}_1/m_1 )
R B(\mathbf{p}_1/m_1 ) \mathbf{j}_1
\label{h.13}
\eeq
This is a different Wigner rotation of R, but it also involves 
rotationless boosts, so I have 
\beq 
B^{-1}(R \mathbf{p}_1/m_1 )
R B(\mathbf{p}_1/m_1 ) = R
\label{h.14}
\eeq
Similar results hold for $\mathbf{j}_2$. 

Since the quantities $\mathbf{k}$, $\mathbf{j}_1$ and 
$\mathbf{j}_2$ all rotate together, they can be coupled 
using ordinary spherical harmonics and $SU(2)$ Clebsch-Gordan 
coefficients:
\[
\vert \mathbf{0},k, l, s, j, \mu \rangle :=
\]
\[
\int  
\vert \mathbf{k} , \mu_1 \rangle \otimes 
\vert -\mathbf{k} , \mu_2 \rangle \times d\Omega (\hat{\mathbf{k}})
Y^{l}_{\mu_l} (\hat{\mathbf{k}})  
\]
\beq 
\langle j_1, \mu_1, j_2, \mu_2 \vert s, \mu_s \rangle 
\langle l \mu_l, j_2, s , \mu_s \vert j ,\mu \rangle 
\label{h.15}
\eeq
This state is a zero momentum eigenstate of state the two 
particle systems that transforms under a $2j+1$-dimensional 
representation of the rotation group.  It satisfies 
\beq
U(R,0) \vert \mathbf{0},k, l, s, j, \mu \rangle :=
\vert \mathbf{0},k, l, s, j, \mu' \rangle
D^j_{\mu'\mu}(R)
\label{h.16}
\eeq

I use the factorization theorem to construct irreducible representations.  
It follows from (\ref{g.8}) that the irreducible state with  
linear momentum $\mathbf{P}$ is    
\[
\vert \mathbf{P},k, l, s, j, \mu \rangle
\]
\beq
U_0 [B^{-1}(\mathbf{P}/M_0)] \vert \mathbf{0},k, l, s, j, \mu \rangle
\sqrt{M_0 \over \sqrt{\mathbf{P}^2+M_0^2}}
\label{h.17}
\eeq
where as in (\ref{g.8}) the multiplicative factor is chosen to make  
$U_0 [B^{-1}(\mathbf{P}/M_0)]$ unitary if
$\vert \mathbf{P},k, l, s, j, \mu \rangle $ has a 
$\delta (\mathbf{P}-\mathbf{P}')$ normalization.

Equation (\ref{h.15}) can be inverted to give
\[
\vert \mathbf{k} , \mu_1 \rangle \otimes 
\vert -\mathbf{k} , \mu_2 \rangle =
\]
\beq
\sum \vert \mathbf{0},k, l, s, j, \mu \rangle  
Y^{l*}_{\mu_l} (\hat{\mathbf{k}}) 
\langle l \mu_l j_2 s  \mu_s \vert j \mu \rangle
\langle j_1 \mu_1 j_2 \mu_2 \vert s \mu_s \rangle
\label{h.18}
\eeq
Using (\ref{h.18}) in (\ref{h.7}) gives
\[
U_0 [B^{-1}(\mathbf{P}/M_0)] 
\vert \mathbf{p}_1 , \mu_1 \rangle \otimes 
\vert \mathbf{p}_2 , \mu_2 \rangle  =
\]
\[
\sum \vert \mathbf{0},k, l, s, j, \mu \rangle  
Y^{l*}_{\mu_l} (\hat{\mathbf{k}}) 
\langle l, \mu_l, s,  \mu_s \vert j ,\mu \rangle
\langle j_1, \mu_1', j_2, \mu_2' \vert s, \mu_s \rangle
\times
\]
\[
\sqrt{\omega_{m_1}(\mathbf{k}) \over 
\omega_{m_1}(\mathbf{p}_1)} 
\sqrt{\omega_{m_2}(\mathbf{k}) \over 
\omega_{m_2}(\mathbf{p}_2)} \times
\]
\[
D^{j_1}_{\mu_1' \mu_1} [B^{-1}(\mathbf{k}/m_1)
B^{-1}(\mathbf{P}/M_0)B (\mathbf{p}_1/m_1)] \times 
\]
\beq
D^{j_2}_{\mu_2' \mu_2} [B^{-1}(-\mathbf{k}/m_2)
B^{-1}(\mathbf{P}/M_0)B (\mathbf{p}_2/m_2)] 
\label{h.19}
\eeq

Combining (\ref{h.17}) and (\ref{h.19}) gives the desired decomposition of
tensor products of irreducible representations into 
\[
\vert \mathbf{p}_1 , \mu_1 \rangle \otimes 
\vert \mathbf{p}_2 , \mu_2 \rangle  =
\]
\[
\sum \vert \mathbf{P},k, l, s, j, \mu \rangle 
\sqrt{M_0 \over \sqrt{\mathbf{P}^2+M_0^2}}
\sqrt{\omega_{m_1}(\mathbf{k}) \over 
\omega_{m_1}(\mathbf{p}_1)} 
\sqrt{\omega_{m_2}(\mathbf{k}) \over 
\omega_{m_2}(\mathbf{p}_2)} \times
\]
\[
 Y^{l*}_{\mu_l} (\hat{\mathbf{k}}) 
\langle l, \mu_l, j_2, s , \mu_s \vert j, \mu \rangle
\langle j_1, \mu_1', j_2, \mu_2' \vert s ,\mu_s \rangle
\times
\]
\[
D^{j_1}_{\mu_1' \mu_1} [B^{-1}(\mathbf{k}/m_1)
B^{-1}(\mathbf{P}/M_0)B(\mathbf{p}_1/m_1)] \times
\]
\beq
D^{j_2}_{\mu_2' \mu_2} [B^{-1}(-\mathbf{k}/m_2)
B^{-1}(\mathbf{P}/M_0)B(\mathbf{p}_2/m_2)] 
\label{h.20}
\eeq
The Clebsch-Gordan coefficients can be read off by taking 
matrix elements with single particle states
{\color{red}
\[
\langle \mathbf{P},k, l, s, j, \mu  \vert 
\mathbf{p}_1 , \mu_1 , \mathbf{p}_2 , \mu_2 
\rangle
=
\]
\[ 
\int 
d \Omega (\hat{\mathbf{k}}) 
\delta (\mathbf{p}_1 - \mathbf{p}_1(\mathbf{P}, \mathbf{k}))
\delta (\mathbf{p}_2 - \mathbf{p}_2(\mathbf{P}, \mathbf{k}))
\times
\]
\[
\sqrt{M_0 \over \sqrt{\mathbf{P}^2+M_0^2}}
\sqrt{\omega_{m_1}(\mathbf{k}) \over 
\omega_{m_1}(\mathbf{p}_1)} 
\sqrt{\omega_{m_2}(\mathbf{k}) \over 
\omega_{m_2}(\mathbf{p}_2)} \times
\]
\[
 Y^{l*}_{\mu_l} (\hat{\mathbf{k}}) 
\langle l, \mu_l, j_2, s,  \mu_s \vert j, \mu \rangle
\langle j_1, \mu_1', j_2, \mu_2' \vert s, \mu_s \rangle
\times
\]
\[
D^{j_1}_{\mu_1' \mu_1} [B^{-1}(\mathbf{k}/m_1)
B^{-1}(\mathbf{P}/M_0)B(\mathbf{p}_1/m_1)]
\times
\]
\beq 
D^{j_2}_{\mu_2' \mu_2} [B^{-1}(-\mathbf{k}/m_2)
B^{-1}(\mathbf{P}/M_0)B(\mathbf{p}_2/m_2)] =
\label{h.21}
\eeq
\[ 
\delta (\mathbf{P} - \mathbf{p}_1- \mathbf{p}_2))
\delta (\mathbf{k} - \mathbf{k}_1(\mathbf{p}_1, \mathbf{p}_2))
\times
\]
\[
\sqrt{\sqrt{\mathbf{P}^2+M_0^2}\over M_0 }
\sqrt{\omega_{m_1}(\mathbf{p}_1) \over 
\omega_{m_1}(\mathbf{k})} 
\sqrt{\omega_{m_2}(\mathbf{p}_2) \over 
\omega_{m_2}(\mathbf{k})} \times
\]
\[
Y^{l*}_{\mu_l} (\hat{\mathbf{k}}) 
\langle l, \mu_l, j_2, s , \mu_s \vert j, \mu \rangle
\langle j_1, \mu_1', j_2, \mu_2' \vert s, \mu_s \rangle
\times
\]
\[
D^{j_1}_{\mu_1' \mu_1} [B^{-1}(\mathbf{k}/m_1)
B^{-1}(\mathbf{P}/M_0)B(\mathbf{p}_1/m_1)] 
\times
\]
\beq
D^{j_2}_{\mu_2' \mu_2} [B^{-1}(-\mathbf{k}/m_2)
B^{-1}(\mathbf{P}/M_0)B(\mathbf{p}_2/m_2)] 
\label{h.22}
\eeq
}
where in the second term I used the relation 
\beq
{\partial (\mathbf{P} \mathbf{k} ) \over 
\partial (\mathbf{p}_1 \mathbf{p}_2)}=
{\sqrt{\mathbf{P}^2+M_0^2}\over M_0 }
{\omega_{m_1}(\mathbf{p}_1) \over 
\omega_{m_1}(\mathbf{k})} 
{\omega_{m_2}(\mathbf{p}_2) \over 
\omega_{m_2}(\mathbf{k})} 
\label{h.23}
\eeq
to change the variables that appear in the delta functions.

The new feature of the Poincar\'e Clebsch-Gordan coefficients that 
do not appear in the corresponding $SU(2)$ Clebsch-Gordan coefficients 
is the appearance of the quantum numbers $l$ and $s$.  Intuitively 
these correspond to the spin and orbital angular momentum.  In the 
Clebsch-Gordan coefficient (\ref{h.21}) or (\ref{h.22}) they 
represent degeneracy parameters, indicating that in decomposing 
the tensor product of two irreducible representations into 
irreducible representations, the same values of $M_0$ and $\mathbf{j}$
appear more than once.   In these expression we have replaced 
$M_0$ by the continuous variable $k:=\vert \mathbf{k} \vert$;
where $k$ and $M_0$ are related by 
\beq
M_0= \sqrt{m_1^2 + \mathbf{k}^2} + \sqrt{m_2^2 + \mathbf{k}^2}
\eeq

In general the structure of the Clebsch-Gordan coefficients depends on
the choice of basis for the irreducible representations.  
In the above I chosen the irreducible basis to
be simultaneous eigenstates of linear momentum and $\hat{\mathbf{z}}$
of the canonical spin.  Different choices of basis are possible and 
used in applications.

The key property of the rotationless boost is the observation that 
the Wigner rotation of a rotation is the rotation.  This justified
the used of the partial wave analysis of the rest vector.

This property of the rotationless boost is elementary 
to prove using the $SL(2,\mathbb{C})$ representations.  First note
that (\ref{d.10}) applied for $A=e^{{i \over 2} \btheta \cdot \bsig}$
associated with a rotation $R$ gives 
\beq
e^{{i \over 2} \btheta \cdot \bsig} x^{\mu} \sigma_{\mu}
e^{-{i \over 2} \btheta \cdot \bsig} =
(Rx)^{\mu} \sigma_{\mu} = x^{\mu} (R^{-1}\sigma)_{\mu}
\label{h.24}
\eeq
Using (\ref{h.24}) in the $SL(2,\mathbb{C})$ representation of the 
Wigner rotation of the rotation $R$ gives 
\[
e^{-{1 \over 2} R \brho \cdot \bsig} e^{{i \over 2} \btheta \cdot \bsig} 
e^{{1 \over 2}  \brho \cdot \bsig} =
e^{{i \over 2} \btheta \cdot \bsig} 
e^{-{i \over 2} \btheta \cdot \bsig}
e^{-{1 \over 2} R \brho \cdot \bsig} e^{{i \over 2} \btheta \cdot \bsig} 
e^{{1 \over 2}  \brho \cdot \bsig} =
\]
\beq
e^{{i \over 2} \btheta \cdot \bsig} 
e^{-{1 \over 2} R \brho \cdot R \bsig}  
e^{{1 \over 2}  \brho \cdot \bsig} =
e^{{i \over 2} \btheta \cdot \bsig} 
\label{h.25}
\eeq
which is the $SL(2,\mathbb{C})$ representation of the original 
rotation.

\section {Dynamical representations}

The representation $U(\Lambda, a)$ is abstract.  Specific realizations
of Poincar\'e invariant quantum mechanics involve both a choice of
Hilbert space representations on which $U(\Lambda ,a)$ acts and a
particular implementation of the dynamics.  The problem of adding
interactions to the Hamiltonian while retaining the Poincar\'e
commutation relations is a non-linear problem.  The problem is most easily understood by considering the commutator
\beq
[K^{i} , P^j]= i \delta_{ij} H  
\label{i.1}
\eeq
implies that one cannot add interactions to the Hamiltonian without 
also adding them to the left side of the commutator.  Once interactions 
appear in the operators on the left side of the commutator one must still
satisfy the commutation relations.  

While this problem is non-linear, the resulting generators must satisfy the 
Poincar\'e Lie algebra, and the resulting unitary representation of the 
Poincar\'e group must have a decomposition into irreducible representations.  
A sensible strategy is to add interactions to a direct integral of 
irreducible representations that does not change the group structure.

I discuss the two-particle case \cite{Ba53}, but the result can be extended 
to any number of particles \cite{Co65}\cite{So77}\cite{Co82}.  The starting point is a description of 
two free particles.  The Hilbert space is the tensor product of two
one-particle irreducible representation spaces
\beq
\vert (m_1,j_1) \mathbf{p}_1,\mu_1 \rangle \otimes  
\vert (m_2,j_2) \mathbf{p}_2,\mu_2 \rangle 
\label{i.2}
\eeq

Using the Clebsh-Gordan coefficients derived in section 8 this can be 
replaced by the irreducible basis 
\beq
\vert (k,j) \mathbf{P},\mu , l ,s  \rangle
\label{i.3}
\eeq
where 
\[
U_0(\Lambda ,a) \vert (k,j) \mathbf{P},\mu , l ,s  \rangle =
\vert (k,j) \bLam {P},\mu' , l ,s  \rangle \times
\]
\beq
\sqrt{{\omega_{M_0} (\bLam P) \over \omega_{M_0} (\mathbf{P} )}}
D^j_{\mu'\mu}[B^{-1}(\bLam P/M_0) \Lambda B(\mathbf{P}/M_0) ]
e^{ i\Lambda P \cdot a}
\label{i.4}
\eeq
Here $U_0(\Lambda ,a)$ indicates the non-interacting unitary 
representation of the Poincar\'e group.

Note that the only variables that are transformed in this basis are
$\mathbf{P}$ and $\mu$.  The noninteracting mass operator in this
representation is  
\beq
M_0= \sqrt{m_1^2 + \mathbf{k}^2} + \sqrt{m_2^2 + \mathbf{k}^2}
\label{i.5}
\eeq

The basis vectors are simultaneous eigenstates
of $\mathbf{M}_0$, $j^2$, $\mathbf{j}\cdot \hat{\mathbf{z}}$,
and $\mathbf{P}$.  All of these operators, in addition to the 
conjugate operators, commute with $M_0$. 
  
Comparing to a non-relativistic model, $M_0$ is the relativistic 
analog of the center of mass Hamiltonian.  Its eigenvalues represent the 
energy of the non-interacting two particle-system in the two-particle
rest frame.  

Following the non-relativistic procedure I add an interaction to 
$M_0$:
\beq
M=M_0+V
\label{i.6}
\eeq
to construct an interacting mass operator.  It is desirable to do
this is a manner that does not disrupt the Poincar\'e commutation
relations.  The simplest way to do this is to require that the
interaction commutes with $j^2$, $\mathbf{j}\cdot \hat{\mathbf{z}}$,
and $\mathbf{P}$ and the operators conjugate to $\mathbf{j}\cdot
\hat{\mathbf{z}}$, and $\mathbf{P}$.  It follows that if this
interaction is evaluated in the non-interacting irreducible basis
then it must have the form
\[
\langle  (k',j') \mathbf{P}',\mu' , l' ,s'  \vert V \vert 
(k,j) \mathbf{P},\mu , l ,s  \rangle =
\]
\beq
\delta (\mathbf{P}' - \mathbf{P})
\delta{j'j} \delta_{\mu'\mu} \langle k',l,'s' \Vert v^j \Vert k,l,s \rangle
\label{i.7}
\eeq
The requirement that $V$ commutes with the operators conjugate to
$\mathbf{j}\cdot \hat{\mathbf{z}}$, and $\mathbf{P}$ means that the
reduced kernel $\langle k',l,'s' \Vert v^j \Vert k,l,s \rangle$ does
not depend on $\mathbf{P}$ or ${\mu}$.  It is encouraging to note that
this interacting has the same number of degrees of freedom
as a rotationally invariant non-relativistic interaction in a partial-wave
basis.

Because $M$ and $M_0$ satisfy the same commutation relations with 
with $j^2$, $\mathbf{j}\cdot \hat{\mathbf{z}}$,
and $\mathbf{P}$ and the operators conjugate to $\mathbf{j}\cdot
\hat{\mathbf{z}}$, and $\mathbf{P}$, it follows that 
it is possible to find simultaneous eigenstates of $M$ 
$j^2$, $\mathbf{j}\cdot \hat{\mathbf{z}}$,
and $\mathbf{P}$, and furthermore that these eigenstates 
transform just like the corresponding non-interacting irreducible 
states, with the eigenvalue of $M_0$ replaced by the eigenvalue of 
$M$.  The desired eigenstates can be constructed by diagonalizing 
$M$ in the free-particle irreducible basis.
\beq
\langle  (k',j') \mathbf{P}',\mu' , l' ,s' 
\vert (\lambda , j) \mathbf{P}',\mu' \rangle =
\eeq
\beq
\delta_{j'j}\delta (\mathbf{P}'-\mathbf{P})  
\delta_{\mu'\mu} 
\langle j', k', l', s' \vert j, \lambda \rangle
\label{i.8}
\eeq
\beq
(\sqrt{m_1^2+ \mathbf{k}^{ 2}} + \sqrt{m_2^2+ \mathbf{k}^{ 2}} )
\langle j, k, l, s \vert j, \lambda \rangle + 
\eeq
\beq
\sum_{s=\vert j_1-j_2 \vert}^{\vert j_1+j_2 \vert}
\sum_{l=\vert j-s \vert}^{\vert j+s \vert} \int_0^{\infty} 
\langle k,l,s \Vert v^j \Vert k',l',s' \rangle k^{\prime 2} dk'
\langle j, k', l', s' \vert j, \lambda \rangle
\eeq
\beq
= \lambda \langle j, k', l', s' \vert j, \lambda \rangle
\label{i.9}
\eeq
For a reasonable interaction these states will be complete.
The commutation relations imply 
\[
U(\Lambda ,a) \vert (\lambda,j) \mathbf{P},\mu  \rangle =
\]
\beq
\vert (\lambda,j) \bLam {P},\mu'  \rangle
\sqrt{{\omega_{\lambda} (\bLam P) \over \omega_{\lambda} (\mathbf{P} )}}
D^j_{\mu'\mu}[B^{-1}(\bLam P/\lambda) \Lambda B(\mathbf{P}/\lambda) ]
e^{ i\Lambda P \cdot a}
\label{i.10}
\eeq
This leads defines the dynamical representation of the Poincar\'e group.

The eigenfunctions and the Clebsch-Gordan coefficients can be used to 
express these relations in plane wave bases.   

This basic construction was first done by Bakamjian and Thomas, \cite{Ba53}.
They used the same basis for the irreducible representation the I used 
above, resulting in an ``instant-form dynamics''\cite{dirac},  
where the mass eigenvalue
$\lambda$ does not appear in the coefficients (\ref{i.10}) when the 
Poincar\'e transformation is a rotation or spatial translation.

\section{Fields}  

While Poincar\'e invariant dynamics is not a local field theory, it is
possible to construct fields that transform covariantly with respect
to the dynamical representation of the Poincar\'e group.  An important
example of a field in applications of this formalism is a current
operator, however it is easy to construct fields of any spin.

The method of construction is based on the Wigner-Eckart theorem for 
the Poincar\'e group.  In general a covariant field is a set of operators
that depend on a space-time coordinate $x$:
\beq
\Psi_n (x)
\label{j.1}
\eeq
and transform covariantly:
\beq
U(\Lambda ,a) \Psi_n (x)
U^{\dagger} (\Lambda ,a) = \Psi_{n'} (\Lambda x + a ) S(\Lambda)_{n'n}
\label{j.2}
\eeq
where $S(\Lambda)_{n'n}$ is a finite-dimensional representation of the
Lorentz group.  These representations are well known; the irreducible 
building blocks are symmetrized tensor products of $SL(2,\mathbb{C})$ 
matrices \cite{St65}.  

There is a large class of operators that satisfy (\ref{j.2}).  A field
operator is defined if all of its matrix elements in a given basis are
known.  I evaluate the matrix elements of field operators 
in the basis of irreducible eigenstates of $U(\Lambda ,a)$.  To specify 
the operator $\Psi_n (x)$ it is necessary to determine the  
matrix elements:
\beq
\langle (m',j') \mathbf{p}' , \mu' \vert 
\Psi_n (x) \vert (m,j) \mathbf{p} , \mu \rangle
\label{j.3}
\eeq
Inserting $U^{\dagger} (\Lambda ,a) U (\Lambda ,a) =I$
on both sides of the field operators in (\ref{j.3}), using 
the covariance relation, implies the identity
\[
\langle (m',j') \mathbf{p}' , \mu' \vert 
\Psi_n (x) \vert (m,j) \mathbf{p} , \mu \rangle =
\]
\[
\langle
(m',j') \bLam p',\nu' \vert
\Psi_{n'} (\Lambda x + a ) \vert 
(m,j) \bLam p,\nu \rangle \times
\]
\[
e^{i \Lambda p \cdot a } 
D^j_{\nu \mu} 
[ B^{-1} (\bLam p/m)
\Lambda B(\mathbf{p}/m) ] 
\sqrt{ {\omega (\bLam p ) \over \omega (\mathbf{ p} )} }\times 
\]
\beq
e^{-i \Lambda p' \cdot a } 
D^{j*}_{\nu' \mu'} 
[ B^{-1} (\bLam p'/m')
\Lambda B(\mathbf{p}'/m') ] 
\sqrt{ {\omega (\bLam p' ) \over \omega (\mathbf{ p}' )} } 
S(\Lambda)_{nm}.
\label{j.4}
\eeq
This equation can be used to relate an arbitrary matrix element
to a reduced set of invariant independent matrix elements.

For example, if I set $\Lambda =I$ and $a=-x$ in (\ref{j.4}) I get
\[
\langle (m',j') \mathbf{p}' , \mu' \vert 
\Psi_n (x) \vert \langle (m,j) \mathbf{p} , \mu \rangle =
\]
\beq
\langle
(m',j') \mathbf{p}',\mu' \vert
\Psi_n (0 ) \vert 
(m,j) \mathbf{p},\mu \rangle
e^{i (p'-p) \cdot x } 
\label{j.5}
\eeq
which shows that matrix elements of the field operator for any $x$ can
be expressed, using translational covariance, in terms of matrix elements 
with $x=0$.

Since the initial four momentum is time-like it is possible to
use a rotationless Lorentz transformation 
$\Lambda = B^{-1} (\mathbf{p}/m)$, $a=0$ to transform 
the initial momentum to its rest value:
\[
\langle (m',j') \mathbf{p}' , \mu' \vert 
\Psi_n (0) \vert (m,j) \mathbf{p} , \mu \rangle =
\]
\[
\langle
(m',j') \bLam p',\nu' \vert
\Psi_{n'} (0 ) \vert 
(m,j) \mathbf{0},\mu \rangle 
\sqrt{ {m \over \omega (\mathbf{ p} )}} 
\sqrt{ {\omega (\bLam p' ) \over \omega (\mathbf{ p}' )}}\times
\]
\beq  
D^{j*}_{\nu' \mu'} 
[ B^{-1} (\bLam p'/m')
\Lambda B(\mathbf{p}'/m') ] 
S(\Lambda)_{n'n}.
\label{j.6}
\eeq
Equation (\ref{j.6}) , along with (\ref{j.5}) 
implies that all matrix elements can be expressed in terms of the 
matrix elements
\beq
\langle (m',j') \mathbf{p}' , \mu' \vert 
\Psi_n (0) \vert (m,j) \mathbf{0} , \mu \rangle . 
\label{j.7}
\eeq
Finally I can use a rotation $R$ about an axis parallel to 
$\hat{\mathbf{z}} \times \mathbf{p}'$ to orient $\mathbf{p}'$ 
in the $\hat{z}$ direction. 
\[
\langle (m',j') \mathbf{p}' , \mu' \vert 
\Psi_n (0) \vert (m,j) \mathbf{0} , \mu \rangle =
\]
\[
\langle
(m',j') \hat{\mathbf{z}}\vert \mathbf{p}' \vert,\
\nu' \vert
\Psi_{n'} (0 ) \vert 
(m,j) \mathbf{0},\nu \rangle \times
\]
\beq 
D^{j}_{\nu \mu} [R] 
D^{j*}_{\nu' \mu'} 
[R ] 
S(R)_{n'n}
\label{j.8}.
\eeq
Combining (\ref{j.5}) with (\ref{j.6}) and (\ref{j.8}) 
implies that every matrix element of $\Psi_n (x)$ can be expressed 
in terms of the matrix element
\beq
\langle
(m',j') \hat{\mathbf{z}}\vert \mathbf{p}' \vert,\
\nu' \vert
\Psi_{n'} (0 ) \vert 
(m,j) \mathbf{0},\nu \rangle .
\label{j.9}
\eeq
Finally I can still use rotations about the z axis to constrain the 
discrete indices
\[ 
\langle
(m',j') \hat{\mathbf{z}}\vert \mathbf{p}' \vert,\
\nu' \vert
\Psi_{n'} (0 ) \vert 
(m,j) \mathbf{0},\nu \rangle =
\]
\beq
\langle
(m',j') \hat{\mathbf{z}}\vert \mathbf{p}' \vert,\
\nu' \vert
\Psi_{n'} (0 ) \vert 
(m,j) \mathbf{0},\nu \rangle e^{i (\mu' -\mu) \phi}S_{n'n} [R(\phi)] . 
\label{j.10}
\eeq
Differentiating with respect to $\phi$ and setting $\phi=0$ gives 
\[
(\mu' -\mu)
\langle
(m',j') \hat{\mathbf{z}}\vert \mathbf{p}' \vert,\
\nu' \vert
\Psi_{n} (0 ) \vert 
(m,j) \mathbf{0},\nu \rangle =
\]
\beq
i \langle
(m',j') \hat{\mathbf{z}}\vert \mathbf{p}' \vert,\
\nu' \vert
\Psi_{n'} (0 ) \vert 
(m,j) \mathbf{0},\nu \rangle
{\partial \over \partial \phi}  S_{n'n} [R(\phi)]_{\vert_{\phi=0}}  
\label{j.11}
\eeq

The last constraint has to be evaluated on a case by case basis.  For
a 4-vector field
\beq
\Psi_{n'} (0 )\to  J^{\alpha}(0)
S_{n'n} [R(\phi)] \to 
\left (
\begin{array}{cccc}
1 & 0 & 0 & 0 \\
0 & \cos (\phi) & \sin (\phi) & 0 \\
0 & -\sin (\phi) & \cos (\phi) & 0 \\
0 & 0 & 0 & 1 \\
\end{array}
\right ) 
\label{j.12}
\eeq
and 
\beq
{\partial \over \partial \phi}  S_{n'n} [R(\phi)]_{\vert_{\phi=0}}  \to 
\left (
\begin{array}{cccc}
0 & 0 & 0 & 0 \\
0 & 0 & 1 & 0 \\
0 & -1 & 0 & 0 \\
0 & 0 & 0 & 0 \\
\end{array}
\right ) 
\label{j.13}
\eeq
which gives the following constraints:

\[
\langle
(m',j') \hat{\mathbf{z}}\vert \mathbf{p}' \vert,\
\nu' \vert
J^0(0)  \vert 
(m,j) \mathbf{0},\nu \rangle =
\]
\beq
\delta_{\nu \nu'}
\langle
(m',j') \hat{\mathbf{z}}\vert \mathbf{p}' \vert,\
\nu \vert
J^0(0)  \vert 
(m,j) \mathbf{0},\nu \rangle
\label{j.14}
\eeq

\[
\langle
(m',j') \hat{\mathbf{z}}\vert \mathbf{p}' \vert,\
\nu' \vert
J^3(0)  \vert 
(m,j) \mathbf{0},\nu \rangle =
\]
\beq
\delta_{\nu \nu'}
\langle
(m',j') \hat{\mathbf{z}}\vert \mathbf{p}' \vert,\
\nu \vert
J^3(0)  \vert 
(m,j) \mathbf{0},\nu \rangle
\label{j.15}
\eeq

\[
(\mu' -\mu) \langle
(m',j') \hat{\mathbf{z}}\vert \mathbf{p}' \vert,\
\nu' \vert
J^x (0 ) \vert 
(m,j) \mathbf{0},\nu \rangle =
\]
\beq
i \langle
(m',j') \hat{\mathbf{z}}\vert \mathbf{p}' \vert,\
\nu' \vert
J^y (0 ) \vert 
(m,j) \mathbf{0},\nu \rangle
\label{j.16}
\eeq

\[
(\mu' -\mu) \langle
(m',j') \hat{\mathbf{z}}\vert \mathbf{p}' \vert,\
\nu' \vert
J^y (0 ) \vert 
(m,j) \mathbf{0},\nu \rangle =
\]
\beq
- i \langle
(m',j') \hat{\mathbf{z}}\vert \mathbf{p}' \vert,\
\nu' \vert
J^x (0 ) \vert 
(m,j) \mathbf{0},\nu \rangle
\label{j.17}
\eeq
The last two of these equation can combined to give
\beq
[(\mu' -\mu)^2 -1] \langle
(m',j') \hat{\mathbf{z}}\vert \mathbf{p}' \vert,\
\nu' \vert
J^x (0 ) \vert 
(m,j) \mathbf{0},\nu \rangle = 0
\label{j.18}
\eeq
\beq
[(\mu' -\mu)^2 -1] \langle
(m',j') \hat{\mathbf{z}}\vert \mathbf{p}' \vert,\
\nu' \vert
J^y (0 ) \vert 
(m,j) \mathbf{0},\nu \rangle = 0
\label{j.19}
\eeq
\[
\langle
(m',j') \hat{\mathbf{z}}\vert \mathbf{p}' \vert,\
\nu' \vert
J^y (0 ) \vert 
(m,j) \mathbf{0},\nu \rangle =
\]
\beq
{  i \over (\mu -\mu') } \langle
(m',j') \hat{\mathbf{z}}\vert \mathbf{p}' \vert,\
\nu' \vert
J^x (0 ) \vert 
(m,j) \mathbf{0},\nu \rangle
\label{j.20}
\eeq

This implies that the most general for vector field, $J^{\mu}(x)$,
can be uniquely specified by defining the independent matrix elements and 
using covariance to generate the remaining matrix elements:
\beq
\langle
(m',j') \hat{\mathbf{z}}\vert \mathbf{p}' \vert,\
\nu \vert
J^0(0)  \vert 
(m,j) \mathbf{0},\nu \rangle 
\label{j.21}
\eeq
\beq
\langle
(m',j') \hat{\mathbf{z}}\vert \mathbf{p}' \vert,\
\nu \vert
J^3(0)  \vert 
(m,j) \mathbf{0},\nu \rangle 
\label{j.22}
\eeq
\beq
\langle
(m',j') \hat{\mathbf{z}}\vert \mathbf{p}' \vert,\
\nu \pm 1  \vert
J^x (0 ) \vert 
(m,j) \mathbf{0},\nu \rangle 
\label{j.23}
\eeq
These independent matrix elements are Poincar\'e invariant functions - 
since one arrives that the same independent matrix elements 
from any starting frame.

These invariant matrix elements are the analog of the reduced 
matrix elements that appear in the standard $SU(2)$ Wigner Eckart theorem.

The form of the Wigner-Eckart theorem does not look exactly like the
standard form because the operators are expressed as Lorentz covariant
densities; however they could {\it equivalently} be expressed as
Poincar\'e covariant operators \cite{wpwk}.  This is not normally done
because as the Fourier transform of $x$ passes through the six classes
of irreducible representation in Table 1, the transformation
properties of the operator change.  With Lorentz covariant densities
transformation properties of the continuous parameter $x$ is decoupled
from the discrete field index.

If $J^{\mu}(x)$ is an electromagnetic current operator then
current conservation and parity will further reduce the number of 
independent invariant matrix elements.  The resulting number 
corresponds exactly to the number of invariant form factors. 

The two main messages are from this section are (1) even though the
representation $U(\Lambda ,a)$ is not manifestly covariant, the theory
has many fields that transform covariantly and (2) covariant field
operators can be constructed using the Wigner-Eckart theorem for the
Poincar\'e group.

\section{Examples}

\subsection{Confined Quarks}

Assume equal mass quarks and antiquarks.

\begin{itemize}

\item [1.] Hilbert Space (identify the degrees of freedom - treat free quarks
as massive spin $1/2$ particles)

\beq
{\cal H} = {\cal H}_q \otimes {\cal H}_{\bar{q}}
\label{k.1}
\eeq

\item [2.] Mass operator (include dynamics - confining interaction)

\beq
M^2 = 4(\mathbf{k}^2 + m^2) - \lambda \nabla_{k}^2 
\label{k.2}
\eeq

\item [3.] Mass eigenfunctions (in the non-interacting irreducible 
basis)

\beq
\langle  (k', j') \mathbf{P}', \mu' ; l', s' \vert
(M,  j) \mathbf{P}, \mu  \rangle =
\delta (\mathbf{P}'- \mathbf{P}) 
\delta_{j'j} \delta_{\mu' \mu} \phi_M^j (k',l',s') 
\label{k.3}
\eeq

\item [4.] Mass eigenvalue problem (solve in the non-interacting irreducible 
basis).

\beq
4(\mathbf{k}^2 +m^2 ) \phi^j (k,l,s)  
- \lambda \bnabla_k \phi^j (k,l,s) = M^2 \phi^j (k,l,s)
\label{k.4}
\eeq

\item [5.] Relativistic Dynamics (simultaneous eigenstates of mass,
spin, linear momentum and z-component of canonical spin are 
complete and transform irreducibly) 
{\color{red}
\[  
\langle (k,j) \mathbf{P}, \mu ; l, s \vert U(\Lambda ,a) \vert
(M,  j) \mathbf{P}, \mu  \rangle =
\]
\[
\langle (k,j) \mathbf{P}, \mu ; l, s \vert U(\Lambda ,a) \vert
(M,j) \bLam {P} , \nu  \rangle \times
\]
\beq 
\sqrt{{\omega_M (\bLam P_M) \over \omega_M (\mathbf{P} )}}
D^j_{\nu \mu}[B^{-1}(\bLam \mathbf{P}/M) \Lambda B (\mathbf{P}/M)]
e^{i \Lambda P_M \cdot a} 
\label{k.5}
\eeq
}
\item [] where
\beq
P_M =( \sqrt{\mathbf{P}^2 +M^2}, \mathbf{P} )
\label{k.6}
\eeq

\item [5.] Representation in terms of quark degrees of freedom (use
Poincar\'e Clebsh-Gordan coefficients - needed to calculate
electromagnetic observables)

\[
\langle \mathbf{p}_q, \mu_q, \mathbf{p}_{\bar{q}} , \mu_{\bar{q}}\vert 
(M,j) \mathbf{P} , \nu  \rangle =
\]
\[
\sqrt{\sqrt{\mathbf{P}^2+M_0^2}\over M_0 }
\sqrt{\omega_{m_q}(\mathbf{p}_q) \over 
\omega_{m_q}(\mathbf{k})} 
\sqrt{\omega_{m_{\bar{q}}}(\mathbf{p}_{\bar{q}}) \over 
\omega_{m_{\bar{q}}}(\mathbf{k})} \times
\]
\[
Y^{l}_{\mu_l} (\hat{\mathbf{k}}) 
\langle l, \mu_l, s  \mu_s \vert j \mu \rangle
\langle j_q ,\mu_q' j_{\bar{q}} \mu_{\bar{q}}' \vert s \mu_s \rangle
\times
\]
\[
D^{j_q*}_{\mu_q' \mu_q} [B^{-1}(\mathbf{k}/m_q)
B^{-1}(\mathbf{P}/M_0)B (\mathbf{p}_q/m_q)] 
\times
\]
\beq
D^{j_{\bar{q}}*}_{\mu_{\bar{q}}' \mu_{\bar{q}}} [B^{-1}(-\mathbf{k}/m_{\bar{q}})
B^{-1}(\mathbf{P}/M_0)B(\mathbf{p}_{\bar{q}}/m_{\bar{q}})] 
\phi^j (k,l,s) 
\label{k.7}
\eeq

\end{itemize} 

These steps show that this model is mathematically equivalent to a
quantum mechanical harmonic oscillator.  Because the oscillator is
associated with the square of the mass, taking square roots, it is
easy to show that the mass eigenvalues grow linearly with the mean
separation of the partons.  More complicated spin-flavor dependent
interactions can be included in step 3.
 
\subsection{Pion production near threshold}

This is an example of a model that does not conserve particle 
number.

\begin{itemize}
\item [1.] Hilbert Space (identify degrees of freedom - in this example 
I treat the pions as physical rather than bare pions.)

\beq
{\cal H} = ({\cal H}_N \otimes {\cal H}_{N})
\oplus ({\cal H}_N \otimes {\cal H}_{N} \otimes {\cal H}_{\pi})
\label{k.8}
\eeq

\item [2.] Mass operator (include dynamics)

\[
M = M_0 + V =
\]
\beq
\left ( 
\begin{array} {cc} 
M_{0NN} & 0 \\
0 & M_{0NN\pi} 
\end{array} 
\right ) 
+
\left ( 
\begin{array} {cc} 
V_{NN} & V_{NN;NN\pi}  \\
V_{NN\pi;NN} & V_{NN}+ V_{N\pi} + V_{N'\pi}    
\end{array} 
\right ) 
\label{k.9}
\eeq

\item [3.] Mass eigenfunction (in the non-interacting irreducible 
basis)

\beq
\left ( 
\begin{array} {c} 
\langle  (k', j') \mathbf{P}', \mu' ; l', s' \vert
(M,  j) \mathbf{P}, \mu  \rangle \\
\langle  (k', q', j') \mathbf{P}', \mu' ; L', S', j_2', l', s' \vert
(M,  j) \mathbf{P}, \mu  \rangle 
\end{array} 
\right ) =
\label{k.10}
\eeq

\beq
\delta (\mathbf{P}' - \mathbf{P})
\delta_{j'j}
\delta_{\mu' \mu}  
\left ( 
\begin{array} {c} 
\phi_{M 1}^j   (k', l', s' ) \\
\phi_{M2}^j   (k', q', L', S', j_2', l', s' ) 
\end{array} 
\right ) 
\label{k.11}
\eeq

\item [4.] Mass eigenvalue problem (in the non-interacting irreducible 
basis).
\[
(M- 2 \sqrt{\mathbf{k}^2 + m_N^2} ) \phi_{M 1}^j   (k', l', s' ) =
\]
\[
\int \sum V^j _{NN} (k, l, s; k', l', s') \phi_{M 1}^j (k', l', s' )
k^{\prime 2} dk' +
\]
\[
\int \sum V^j _{NN;NN\pi} (k, l, s; k',q' ,L',S',j_2', l', s') 
\times
\]
\beq
\phi_{M2}^j   (k', q', L' S' j_2', l', s' ) k^{\prime 2} dk' q^{\prime 2} dq'
\label{k.12}
\eeq

\[
( M- \sqrt{4\mathbf{k}^2 + 4m_N^2 + q^2} + \sqrt{\mathbf{q}^2 + m_\pi^2})
\phi_{M2}^j   (k', q', L' S' j_2'; l', s' )  =
\]
\[
\int \sum V^j _{NN\pi;NN} (k,q,L,S,j_2,l,s; k', l', s') \phi_{M 1}^j (k', l', s' )
k^{\prime 2} dk' +
\]
\[
\int \sum V^j _{NN\pi;NN\pi} (k,q,L,S,j_2,l,s; k',q' ,L',S',j_2', l', s') 
\times
\]
\beq
\phi_{M2}^j   (k', q', L', S', j_2', l', s' ) k^{\prime 2} dk' q^{\prime 2} dq'
\label{k.13}
\eeq
where $V^j_{NN\pi;NN\pi}$ is a sum of two body interactions in
(\ref{k.9}).  Different orders of coupling of the irreducible
representation are natural for each pairwise interaction in the
three-particle sector.  The transformations that change the order of
the coupling can be computed using four Clebsch-Gordan coefficients -
they are the analog of ``Racah coefficients'' for the Poincar\'e
group. 

\item [] The solution to this problem has the complexity of a three-body 
problem.  The scattering problem with all of the correct 
asymptotic conditions must be solved using Faddeev methods. 

\item []  For the pion to be physical the off diagonal parts of the 
interaction should be short ranged ``2-3'' operators rather 
than elementary vertices. 

\item [5.] Relativistic Dynamics (simultaneous eigenstates of mass,
spin, linear momentum and z-component of canonical spin are 
complete and transform irreducibly): 

\[
\left ( 
\begin{array} {c} 
\langle  (k', j') \mathbf{P}', \mu' ; l', s' \vert
U(\Lambda ,a) \vert
(M,  j) \mathbf{P}, \mu  \rangle \\
\langle  (k', q', j') \mathbf{P}', \mu' ; L', S', j_2', l', s' \vert
U(\Lambda ,a) \vert (M,  j) \mathbf{P}, \mu  \rangle 
\end{array} 
\right ) =
\]
\[
\left ( 
\begin{array} {c} 
\langle  (k', j') \mathbf{P}', \mu' ; l', s' \vert
(M,  j) \bLam {P}, \mu'  \rangle \\
\langle  (k', q', j') \mathbf{P}', \mu' ; L', S', j_2', l', s' \vert
(M,  j) \bLam {P}, \mu'  \rangle 
\end{array} 
\right ) \times
\]
\beq
\sqrt{{\omega_M (\bLam P_M) \over \omega_M (\mathbf{P} )}}
D^j_{\mu' \mu}[B^{-1}(\bLam \mathbf{P}/M) \Lambda B (\mathbf{P}/M)]
e^{i \Lambda P_M \cdot a} 
\label{k.14}
\eeq

\item [] where
\beq
P_M =( \sqrt{\mathbf{P}^2 +M^2}, \mathbf{P} )
\label{k.15}
\eeq

\item [5.] Representation in terms of single particle degrees of
  freedom.  The irreducible free-particle basis is constructed by
  first coupling the nucleon irreducible representations to
  two-nucleon irreducible representations and then coupling the
  resulting two-nucleon irreducible representations to the pion
  irreducible representation.  This involves using two sets of
  Poincar\'e Clebsch-Gordan coefficients.  These can be inverted using
  the Poincare Clebsch-Gordan coefficients to express the irreducible
  three particle basis in terms of the tensor product of the single
  particle irreducible bases.  The computation is straightforward, but
  the result is messy and not very illuminating.

\end{itemize} 

This is a simple extension of the NN model to allow for the production
of a single pion.  In this model the deuteron will have a two-nucleon
and two-nucleon one pion component.

\subsection{Quark string model}

The flexibility of Poincar\'e invariant quantum mechanics can be
illustrated by considering a quark model motivated by strong-coupling
lattice QCD.  The basic building blocks of strong coupling Lattice QCD
are quarks and links\cite{Wi75}.  The physical degrees of freedom involve
combinations of quarks, antiquarks and links that are connected to
form a color singlet at each lattice site.  In the absence of the
interactions the energy of each configuration is the sum of the quark
and antiquark masses and a quantity proportional to the total length
of the links, and color singlets are confined.  These degrees of
freedom then interact in a manner that couples color singlets to color
singlets.

To make a Poincar\'e invariant model based on these degrees of freedom
I consider a model with quarks and antiquark degrees of freedom, where
interactions are added in a two step process.  First quarks and
antiquarks connected by links are modeled by quarks and antiquarks
interacting via a confining interaction.  Because this represents a
locally gauge invariant object the color degrees of freedom are
assumed to be summed out.  The mass operator can be diagonalized and
the resulting eigenstates transform like particles.  The interactions
that couple singlets or multi-singlets are modeled by short-range
interactions.  For example, a string breaking interaction is one that
couples one confined singlet quark antiquark pair to a pair of confined
singlet quark-antiquark pairs.  The structure of the interactions can
be motivated by lattice degrees of freedom \cite{Wi75} or axiomatic models 
\cite{Seiler} of
these degrees of freedom.  The interactions between different singlets
can be expressed in terms of internal quark degrees of freedom or the
Poincar\'e irreducible labels of the confined states.

I consider an example of a model that has a meson spectrum, meson 
decay, and meson-meson scattering.  I begin by constructing the 
singlet subspaces.  For this model I include three such subspaces.
The quarks and antiquarks in each subspace are assumed to couple
to color singlets so the quarks in these subspaces do not
have color quantum numbers.  These are tensor products of two or
four irreducible representations of the Poincar\'e group for
each flavor combination:   
\beq
{\cal H}_{q\bar{q}}
\label{k.16}
\eeq
\beq
{\cal H}_{q\bar{q}*}
\label{k.17}
\eeq
\beq
{\cal H}_{qq\bar{q}\bar{q}}
\label{k.18}
\eeq
Each of these subspaces can be decomposed into free particle
irreducible subspaces using the Poincar\'e Clebsh-Gordan coefficients.
An invariant mass operator is defined by adding a confining interaction 
to each non-interacting mass operator:
\beq
M_{q\bar{q}} = M_{0q\bar{q}} + V_{q\bar{q}}  
\label{k.19}
\eeq
\beq
M_{q\bar{q}*} = M_{0q\bar{q}} + V_{q\bar{q}*}  
\label{k.20}
\eeq
\beq
M_{qq\bar{q}\bar{q}} = M_{qq\bar{q}\bar{q}} + V_{qq\bar{q}\bar{q}}  
\label{k.21}
\eeq

It is useful to think of the lattice counterpart of these three
operators in a Born Oppenheimer type of approximation.  For
$V_{q\bar{q}}$ can be thought of as the lowest energy state of a quark
anti-quark pair coupled to a single as a function of the distance
between the quark and antiquark; $V_{q\bar{q}*}$ can be thought of as
the energy of the first excited state of a quark anti-quark pair
coupled to a singlet as a function of the distance between the quark
and antiquark; $V_{qq\bar{q}\bar{q}}$ can be considered as the lowest
energy state of a two quark two anti-quark singlet that cannot be
decomposed into a pair of non-interacting singlets as a function of
the quark and antiquark coordinates.  In Poincar\'e invariant quantum
mechanics these interactions can be modeled.  The interaction in each
of these singlet mass operators is assumed to have only discrete
spectra.

A model Hilbert space is defined by:
\[
{\cal H} = 
\]
\beq
{\cal H}_{q\bar{q}} \oplus 
{\cal H}_{q\bar{q}*} \oplus 
{\cal H}_{qq\bar{q}\bar{q}} \oplus 
({\cal H}_{q\bar{q}} \otimes
{\cal H}_{q\bar{q}} ) \oplus
({\cal H}_{q\bar{q}} \otimes
{\cal H}_{q\bar{q}*} ) \oplus
({\cal H}_{q\bar{q}*} \otimes
{\cal H}_{q\bar{q}*} ) 
\label{k.22}
\eeq

On this space I define the a mass operator that only 
includes interactions between quarks and antiquarks in the 
same color singlet:
\beq
M_c= 
\left (
\begin{array}{ccccccc}
M_{q\bar{q}} & 0 & 0 & 0 & 0 & 0 \\
0 &  M_{q\bar{q}*} & 0 & 0 & 0 & 0 \\
0 &  0 & M_{qq\bar{q}\bar{q}} & 0 & 0 & 0 \\
0 &  0 & 0 & M_{(q\bar{q})(q\bar{q})}  & 0 & 0 \\
0 &  0 & 0 & 0 & M_{(q\bar{q})(q\bar{q})*} & 0 \\
0 &  0 & 0 & 0 & 0 & M_{(q\bar{q}*)(q\bar{q})*}  \\
\end{array} 
\right ) 
\label{k.23}
\eeq
where
\beq
M_{(q\bar{q})(q\bar{q})*} = \sqrt{ M_{(q\bar{q})}^2 + \mathbf{k}^2} +
\sqrt{ M_{(q\bar{q}*)}^2 + \mathbf{k}^2} 
\label{k.24}
\eeq
\beq
M_{(q\bar{q})(q\bar{q})} = \sqrt{ M_{(q\bar{q})_1}^2 + \mathbf{k}^2} +
\sqrt{ M_{(q\bar{q}*)_2}^2 + \mathbf{k}^2} 
\label{k.25}
\eeq
\beq
M_{(q\bar{q})(q\bar{q})*} = \sqrt{ M_{(q\bar{q})_1}^2 + \mathbf{k}^2} +
\sqrt{ M_{(q\bar{q}*)_2}^2 + \mathbf{k}^2} 
\label{k.26}
\eeq
The relative momenta $\mathbf{k}$ in the two-singlet subspaces is 
obtained by using (\ref{h.4}) with $M_0$ replaced by 
one of the above masses, which can also be expressed in terms
of the individual singlet mass eigenvalues and momenta.

Additional interactions allow the quarks in different singlets to
interact.   These interaction have the matrix form 
\beq
\left (
\begin{array}{ccccccc}
0            & V_a &  V_b & V_c & V_d & V_e  \\
V^{\dagger}_a &  0  & V_f & V_g & V_h & V_i  \\
V^{\dagger}_b &  V^{\dagger}_f  & 0 & V_j & V_k & V_l \\
V^{\dagger}_c &  V^{\dagger}_g  & V^{\dagger}_j & V_{s1} & V_m & V_n  \\
V^{\dagger}_d &  V^{\dagger}_h  & V^{\dagger}_k & V^{\dagger}_m & V_{s2} & V+_o \\
V^{\dagger}_e &  V^{\dagger}_i  & V^{\dagger}_l & V^{\dagger}_n & V^{\dagger}_o 
& V_{s3} \\
\end{array} 
\right ) 
\label{k.27}
\eeq
This construction requires that each of interactions commutes with and
is independent of the total $\mathbf{P}$, $\mathbf{j}^2$,
$\mathbf{j}\cdot\hat{\mathbf{z}}$.  In this case the spins are not the
kinematic spins - they are the spins obtained by treating the confined
bound states as particles.  These interactions are assumed to be 
short range interactions when they couple different confined 
irreducible singlets,  however the existence of infinite towers of 
confined singlet states in each sector puts additional constraints
on a model if one wants of have non-trivial meson-meson scattering 
theory. 

Thus, the dynamics of this model is constructed in two steps.  First
the confining mass operator $M_c= M_0+V_c$ is diagonalized in the
non-interacting irreducible basis, to construct a complete set of
confined irreducible mass eigenstates.  Then the $M= M_c +V$ is
diagonalized in the confined irreducible basis to get a dynamical mass
operator, which along with linear momentum and the spin of the
confined system can be used to construct a dynamical irreducible
representation.

This model is a fully relativistic quantum mechanical model.  While
this model it is not complete, it illustrates some of the problems
that need to be addressed in complex systems.  For suitable
interactions this mass operator will support bound states, unstable
resonances, and scattering states.  The bound states correspond to
physical mesons in this model.  These in general will have a different
mass than the corresponding bare mesons that only include the
confining interaction.  The continuous spectrum is associated with
scattering of the bare mesons.  Finally bare mesons with mass in the
scattering continuum should to be unstable.  There is the potential
for an interesting interplay between the resonances and the scattering
states.  The scattering problem involves the sums over an infinite
number of short range interactions - it is not automatic that the sum
of an infinite number of short ranged interactions results in a short
ranged interaction.  This is also related to the decay widths of the
high lying bare mesons.  There will clearly be an interplay between
the lifetimes of high lying states and the existence of a scattering
theory that needs to be investigated in such a model. 
Some of these questions are addressed in \cite{dashen}

\section{\bf Position in Poincar\'e Invariant Quantum Mechanics}

One principle that is given up in Poincar\'e invariant quantum
mechanics is microscopic locality.  The other thing that is given up
is the use of local field operators, which are replaced by particle
degrees of freedom.  Locality cannot be
tested in these theories because there are no suitable position
operators for particles in Poincar\'e invariant quantum theory.  Thus,
while the theory gives up microscopic locality, it also eliminates the
degrees of freedom that are needed to test microscopic locality.

In order to understand the difficulties associated with 
finding a suitable position operator in Poincar\'e invariant 
quantum mechanics 
I begin by considering the wave function of a spinless particle at the
origin at time $t=0$. I denote the wave function of this particle by
$\langle \mathbf{p} | \mathbf{x}=0; t=0 \rangle$.  If I make the 
naively sensible assumption that
{\color{red} such
a state is invariant under homogeneous Lorentz transformations},
then
\[
\langle \mathbf{p} | \mathbf{x}=0; t=0 \rangle
=\langle \mathbf{p} | U(\Lambda ,0 ) | \mathbf{x}=0; t=0 \rangle
=
\]
\beq
 \sqrt{{\omega_m ( \bLam^{-1} {p}  ) \over 
\omega_m ( \mathbf{p} )}} 
\langle \bLam^{-1}{p} | \mathbf{x}=0; t=0 \rangle.
\label{l.1}
\eeq 
Comparing the left and right sides of (\ref{l.1}) it follows that
wave function $\langle \mathbf{p} | \mathbf{x}=0; t=0 \rangle$ 
must have the form
\beq
\langle \mathbf{p} | \mathbf{x}=0; t=0 \rangle = {1 \over \sqrt{\omega_m
(\mathbf{p} )}} f( p^2) = {1 \over \sqrt{\omega_m (\mathbf{p} )}}f (m^2 ) =
{C \over \sqrt{\omega_m (\mathbf{p} )}}, 
\label{l.2}
\eeq
where $C$ is constant. I can now translate 
this eigenstate to construct an eigenstate corresponding to a particle
localized at $\mathbf{x}$:
\beq
\langle \mathbf{p} | \mathbf{x}; t=0 \rangle = \langle \mathbf{p} | 
U(I,\mathbf{x} ) | \mathbf{x}=0; t=0 \rangle = e^{ i \mathbf{p} \cdot \mathbf{x} }
{C \over \sqrt{\omega_m (\mathbf{p} )}}.
\label{l.3}
\eeq
If I take the overlap between the state at $(\mathbf{x}=0;t=0)$
with a state at $(\mathbf{x}\not= 0;t=0)$, the result is (\cite{Bo59}:
\[
\langle \mathbf{0} | \mathbf{x} \rangle = | C | ^2 \int {d^3 p
\over \omega_m (\mathbf{p})} e^{ i \mathbf{p} \cdot \mathbf{x} } 
\]
\[
- (2\pi )^3 | C | ^2 {i \over 2} D_+^* ( 0,\mathbf{x}) =
\]
\beq
 (2\pi )^3 | C | ^2 {i \over 2} \left[ \lim_{t\to 0} {1 \over
4\pi} \epsilon (t) \delta (\mathbf{x}^2 ) - {m i \over 4 \pi^2 | 
\mathbf{x} | } K_1 (m | \mathbf{x} | )\right]  ,
\label{l.4}
\eeq
where $D_+(x)$ is the positive frequency part of the Pauli-Jordan
commutator function.  For $\mathbf{x} \not= 0$, this expression is
non-zero, but falls off like $ K_1 (m | \mathbf{x} | )$, vanishing as
$\vert m \mathbf{x} \vert^{-1/2} e^{- \vert m \mathbf{x} \vert }$ as $
| \mathbf{x} | \to \infty$. Thus, these two states have an overlap
which falls off exponentially when the coordinates are separated by
more than a Compton wavelength. The assumption that a particle
localized at the origin can be described in an invariant way implies
that it is not orthogonal to a state at a different point at the same
time. The Compton wavelength of the particle again sets the scale for
the violation of orthogonality.

It is possible to obtain additional insight by noting that the
position operator discussed above canonically conjugate to the linear
momentum in an irreducible representation. If the representation has a
spin, this operator is also required to commute with canonical spin (
the resulting operator is the so called Newton-Wigner position
operator). This means that it is essentially $-i$ multiplied by the
partial derivative of the linear momentum {\color{red}holding the
canonical spin constant}.  Because the spins undergo momentum
dependent Wigner rotations, this is a non-trivial requirement that
depends on which boost is used to define the spin.

Changing spin observable involves momentum-dependent rotations, which 
do not commute with the momentum derivatives.
To understand this,
consider two spin-$1/2$ wave functions in a canonical spin and helicity
spin basis, respectively:
\beq
_c \langle m \, j ; \mathbf{p} \, \mu | \phi \rangle =
f_{\mu} (\mathbf{p});
\label{l.5}
\eeq
\beq
_h \langle m \, j ; \mathbf{p} \, \mu | \psi \rangle =
f_{\mu} (\mathbf{p}).
\label{l.6}
\eeq
The wave function $f_{\mu} (\mathbf{p})$ is chosen to be the same in each 
case, however the states are different. In both
expressions, $\mathbf{p}$ is the three-momentum. The Fourier transforms
of each of these wave functions are clearly the same. On the
other hand, if we take the wave function 
$_c \langle m \, s ; \mathbf{p} \, \mu | \phi \rangle$, and perform the unitary
transformation that puts it into the same representation as the
wave function $_h \langle m \, s ; \mathbf{p} \, \mu | \psi \rangle$, 
then the new wave function is 
\beq
f_{\mu} (\mathbf{p}) \to f_{\mu}' (\mathbf{p}) = \sum_{\bar\mu}
D^{1\over2}_{\mu {\bar\mu}} \lbrack {R}_{hc} (\mathbf{p}/m )\rbrack
f_{{\bar\mu}} (\mathbf{p} ).
\label{l.7}
\eeq
where
\beq
{R}_{hc}(\mathbf{p}/m)  = B_h^{-1} (\mathbf{p}/m) B_c (\mathbf{p}/m)
\label{l.8}
\eeq
is the momentum dependent rotation constructed using a canonical boost to
from the rest frame to a frame with linear momentum $\mathbf{p}$ followed
by a helicity boost back to the rest frame.  The resulting transformation 
is a momentum dependent rotation, called a generalized Melosh rotation
\cite{Me74}\cite{keipo}.  Obviously the partial derivative with respect to the 
linear momentum holding the canonical spin constant differs from
the partial derivative with respect to the 
linear momentum holding the helicity spin constant.

The conclusion is that although configuration space wave functions can
be used as well as momentum space wave functions, one should never
attempt to interpret the coordinates as observable quantities,
especially on distance scales on the order of a Compton wavelength of
a particle. We note that the concept of position gets even more
complicated in models with interaction dependent spins.

\section{Two-component spinor conventions:}

The group $SL(2,\mathbb{C})$, which is a double cover of the 
Lorentz group, is useful for both computational and
proving simple results \cite{St65}\cite{Wi60}\cite{keipo}. 

\medskip

Let $\sigma_{\mu}$ denote the $2 \times 2$ Pauli spin matrices and the
identity. Let 
\beq
X := x^{\mu} \sigma_{\mu} \qquad x^{\mu} = {1 \over 2} \mbox{Tr} (\sigma_{\mu} X) 
\label{m.1}
\eeq
Note that
\beq
\det (X) = (x^0)^2 - (\mathbf{x}\,)^2 = - x^2 \qquad X= X^{\dagger} 
\label{m.2}
\eeq

Any linear transformation that preserves the determinant and 
Hermiticity of $X$
defines a real Lorentz
transformation.  If $A$ is an arbitrary complex matrix with
$\det (A) =1$ and
\beq
X \to X' = AXA^{\dagger}
\label{m.3}
\eeq
then 
\beq
\det (X')= \det (X) \qquad  \mbox{and} \qquad 
(X')^{\dagger} = X'
\label{m.4}
\eeq

The corresponding Lorentz transformation is
\beq
\Lambda (A)^{\mu}{}_{\nu} = 
{1 \over 2 } \mbox{Tr} (\sigma_{\mu} A \sigma_{\nu} A^{\dagger} )   
\label{m.5}
\eeq
It is obvious from (\ref{m.5}) that 
\beq
\Lambda (A)^{\mu}{}_{\nu} = \Lambda (-A)^{\mu}{}_{\nu}
\label{m.6}
\eeq
It is not difficult to show that there is a 2 to 1 correspondence 
between the $SL(2,\mathbb{C})$ matrices $A$ and Lorentz transformations
connected to the identity.
 
A general element in $SL(2,\mathbb{C})$ has the form
\beq
A = \pm e^{{1 \over 2} ( \brho + i \btheta) \cdot \bsig} .
\label{m.7}
\eeq
\beq
A = e^{{1 \over 2}  \brho \cdot \bsig}
\label{m.8}
\eeq
corresponds to a rotationless boost with rapidity $\brho$ and 
\beq
A = e^{i{1 \over 2}  \btheta \cdot \bsig}
\label{m.9}
\eeq
corresponds to a rotation through an angle $\btheta$.

\noindent {\bf Canonical Boosts:} 

$SL(2,C)$ representatives of canonical boosts are given by:

\beq
\sinh (\omega) = {\vert \mathbf{p} \vert \over m} = \vert \mathbf{v} \vert
\label{m.10}
\eeq
\beq
\cosh (\omega) = {p^0  \over m} = v^0
\label{m.11}
\eeq
\beq
\sinh ({\omega \over 2} ) =  \sqrt{p^0 -m \over 2m}=\sqrt{v^0 -1 \over 2}
\label{m.12}
\eeq
\beq
\cosh ({\omega \over 2} ) =  \sqrt{p^0 +m \over 2m}=\sqrt{v^0 +1 \over 2}
\label{m.13}
\eeq
\[
\Lambda_c(v) := 
\cosh (\omega/2) \sigma_0 + \sinh (\omega/2) \hat{\mathbf{v}} \cdot {\bsig} =
\]
\beq
\sqrt{v^0 +1 \over 2}\sigma_0 + 
\sqrt{v^0 -1 \over 2}\hat{\mathbf{v}} \cdot {\bsig} =
\label{m.14}
\eeq

\beq
{1 \over \sqrt{2(v^0+1)}} \left ( (v^0+1) \sigma_0 + \mathbf{v} 
\cdot {\bsig}\right ) =
\label{m.15}
\eeq

\beq
{1 \over \sqrt{2m(p^0+m)}} \left ( (p^0+m) \sigma_0 + \mathbf{p} 
\cdot {\bsig}\right ) 
\label{m.16}
\eeq

\beq
\Lambda^{\dagger}_c(v)= \Lambda_c(v)
\label{m.17}
\eeq
\beq
\Lambda^{-1}_c (v) = \tilde{\Lambda}_c (v) = 
\cosh (\omega/2) \sigma_0 - \sinh (\omega_2) \hat{\mathbf{v}} \cdot {\bsig} =
\label{m.18}
\eeq
\beq
\sqrt{v^0 +1 \over 2}\sigma_0 -
\sqrt{v^0 -1 \over 2}\hat{\mathbf{v}} \cdot {\bsig} =
\label{m.19}
\eeq
\beq
{1 \over \sqrt{2(v^0+1)}} \left ( (v^0+1) \sigma_0 - \mathbf{v} 
\cdot {\bsig}\right ) =
\label{m.20}
\eeq
\beq
{1 \over \sqrt{2m(p^0+m)}} \left ( (p^0+m) \sigma_0 - \mathbf{p} 
\cdot {\bsig}\right ) 
\label{m.21}
\eeq
Note that in all of the above expressions for the boosts $v^0$ or $p^0$ 
represent on-shell quantities.

\section{Scattering theory in Poincar\'e invariant quantum mechanics} 

\noindent{\bf S and T operators}

The formulation of scattering problems in Poincar\'e invariant quantum
mechanics is based on standard time-dependent multichannel scattering.
Scattering channels $\alpha$ are associated with asymptotically
separated clusters, where the particles in each cluster are either in
a bound state or the cluster consists of a single particle.  Each
distinct partition of the particles into clusters may correspond to
more than one scattering channel or it may support no scattering
channels.  The scattering matrix is the inner product of incoming and
outgoing wave scattering states
\begin{equation}
S_{\alpha \beta} = \langle \Psi_\alpha^+ (0) \vert \Psi_\beta^- (0)
\rangle
\label{ap2.1}
\end{equation}
where $\alpha$ and $\beta$ are channel labels and the initial
and final scattering states are solutions of the time-dependent
Schr\"odinger equation 
\beq
i{d \over dt} \vert \Psi \rangle = H \vert \Psi \rangle
\label{ap2.2}
\eeq
satisfying the incoming and outgoing wave asymptotic conditions
\begin{equation}
\lim_{t \to \pm \infty} \Vert e^{-iHt} \vert \Psi^{\pm}_\alpha (0) \rangle
- \Pi_{\alpha} e^{-i H_\alpha t} \vert \Phi^{\pm}_\alpha (0) \rangle \Vert =0 .
\label{ap2.3}
\end{equation}
In this paper the $\pm$ on the scattering states and wave operators
indicate the direction of the time limit ($-=$past/$+=$future), which
is opposite to the sign of the $i\epsilon$ that appears in the
resolvents used in time independent scattering.  The operator
$\Pi_{\alpha}$ is a channel projection operator,
\beq
\Pi_{\alpha} = \otimes_{i\in \alpha} \sumint_{\nu_i}  
\vert (m_i ,j_i) \mathbf{p}_i, \nu_i \rangle d \mathbf{p}_i
\langle  (m_i ,j_i) \mathbf{p}_i, \nu_i \vert
\label{ap2.4}
\eeq
which projects on the subspace associated with mutually
non-interacting bound subsystems.  The factors $\vert (m_i ,j_i)
\mathbf{p}_i, \nu_i \rangle$ are basis functions for the irreducible
representation of the Poincar\'e group associated with the $i$-th
cluster of the channel $\alpha$ with discrete mass eigenvalue, $m_i$.

The quantity $\vert S_{\alpha \beta}\vert^2$ represents the
probability of a system prepared in a state that in the distant past
looks like a system of asymptotically separated particles in channel
$\beta$ to be measured to be in a state that looks in the asymptotic
future like a system of asymptotically separated particles in channel
$\alpha$.

It follows from (\ref{ap2.3}) that the interacting
and limiting  non-interacting 
asymptotic states are related by the multichannel wave operators
\begin{equation}
\vert \Psi_\alpha^{\pm} (0) \rangle =
\Omega_{\alpha\pm }   (H, H_\alpha)
\vert \Phi_\alpha^{\pm} (0) \rangle
\label{ap2.5}
\end{equation}
where the multichannel wave operators are defined by the strong limits
\begin{equation}
\Omega_{\alpha \pm} = \lim_{t \to \pm \infty}
e^{i H t }\Pi_{\alpha} e^{-i H_\alpha t}.
\label{ap2.6}
\end{equation}
In these equations $H_\alpha$ is the Hamiltonian with the interactions
between particles in different asymptotic clusters set to zero.  The
multichannel scattering operator can then be expressed in terms of the
wave operators as
\begin{equation}
S_{\alpha \beta} =
\Omega_{\alpha + } ^{\dagger}  (H, H_\alpha) \Omega_{\beta -}
(H, H_\beta) .
\label{ap2.7}
\end{equation}
The wave operators can be expressed directly in terms of the mass
operators.  The Kato-Birman invariance principle \cite{Ch76}\cite{baum}
implies that $H$ and $H_\alpha$ in the channel wave operators can be
replaced by $f(H)$ and $f(H_{\alpha})$ where $f$ is any piecewise
differentiable function of bounded variation with positive derivative;
specifically
\begin{equation}
M= \sqrt{H^2 - \mathbf{P}^2}
\label{ap2.8}
\end{equation}
is a function with these properties.  It follows that
\begin{equation}
\Omega_{\alpha \pm} = \lim_{t \to \pm \infty}
e^{i M t }\Pi_{\alpha} e^{-i M_\alpha t} =
\lim_{t \to \pm \infty}
e^{i H t }\Pi_{\alpha} e^{-i H_\alpha t}
\label{ap2.9}
\end{equation}
which leads to the equivalent expression for the
multichannel scattering operator \cite{Co82}:
\begin{equation}
S_{\alpha \beta} = \lim_{\tau,\tau' \to \infty}
e^{i M_\alpha \tau} \Pi_{\alpha}
e^{-iM (\tau+\tau')}\Pi_{\beta}
e^{i M_\beta \tau'} .
\label{ap2.10}
\end{equation}
To relate this to the time-independent formulation of scattering these 
limits are computed in eigenstates $\vert \alpha \rangle$ and
$\vert \beta \rangle$ of
$M_{\alpha}$ and $M_{\beta}$ respectively.  
I prove that
\begin{equation}
\langle \alpha  \vert S \vert \beta \rangle =
\langle \alpha \vert \beta \rangle 
- 2 \pi i \delta ({\sf W}_{\alpha}- {\sf W}_{\beta})
\langle \alpha \vert T^{\alpha \beta}({\sf W}_\alpha +i0^+) \vert
\beta \rangle
\label{ap2.11}
\end{equation}
where
\begin{equation}
T^{\alpha \beta} (z) = V^{\beta } +
V^{\alpha } (z-M)^{-1} V^{\beta} ,
\label{ap2.12}
\end{equation}
and
\begin{equation}
V^{\alpha} = M - M_{\alpha}  
\label{ap2.13}
\end{equation}
and 
\begin{equation}
V^{\alpha} = M - M_{0}  
\label{ap2.14}
\end{equation}
for the breakup channel.

Here ${\sf W}_\alpha$ and ${\sf W}_\beta$ are the eigenvalues
of $M_{\alpha}$ and $M_{\beta}$ in
the channel eigenstates $\vert \alpha \rangle$ and $\vert \beta \rangle$.
The first term in Eq.~(\ref{ap2.11}) is identically zero 
if the states $\vert \alpha
\rangle$ and $\vert \beta \rangle$ correspond to different scattering 
channels.

To prove (\ref{ap2.11}) I evaluate the  
$S$-matrix elements in the channel mass eigenstates:
\begin{eqnarray}
\langle \beta \vert S_{ba} \vert \alpha \rangle 
&=&  \lim_{\tau \to \infty} \langle \beta \vert
e^{i M_\beta \tau} e^{-2iM\tau}e^{iM_\alpha \tau} \vert \alpha\rangle
\nonumber \\
&= &
\langle \beta \vert \alpha \rangle + \lim_{\tau \to \infty}
\int_0^\tau   d\tau'\, {d \over d\tau'}
\langle \beta \vert e^{i({\sf W}_\beta + {\sf W}_\alpha -2M )\tau'}
\vert \alpha \rangle
\nonumber \\
&=&
\langle \beta \vert \alpha \rangle +
\lim_{\epsilon  \to 0^+} i \int_0^\infty   d\tau' \,
\langle \beta \vert 
\nonumber \\
&\times&
\left [ ({\sf W}_\beta -M)e^{i({\sf W}_\beta + {\sf W}_\alpha
-2M+ i\epsilon)\tau'} 
+ e^{i({\sf W}_\beta + {\sf W}_\alpha -2M + i \epsilon )\tau'}
({\sf W}_\alpha -M) \right ]
\vert \alpha \rangle
\nonumber \\
&= &
\langle \beta  \vert \alpha \rangle +
\lim_{\epsilon  \to 0^+}  {1 \over 2}  \langle \beta \vert 
\nonumber \\
&\times&
\left [
(M-{\sf W}_\beta) { 1 \over \bar{{\sf W}} - M + i\epsilon } +
{1 \over \bar{\sf W} - M +
i\epsilon } (M-{\sf W}_\alpha) \right ] \vert \alpha \rangle ,
\label{ap2.15}
\end{eqnarray}

where $M_\alpha \vert \alpha \rangle = {\sf W}_{\alpha} \vert \alpha
\rangle$ and $M_\beta \vert \beta \rangle = {\sf W}_{\beta} \vert
\alpha \rangle$ and $\bar{\sf W} := {1 \over 2} ({\sf W}_\alpha+{\sf
  M}_\beta)$ is the average invariant mass eigenvalues of the initial
and final asymptotic states. The $i\epsilon$ are introduced because
formally the sharp eigenstates should be first integrated against wave
packets in the cluster momenta before the time limit is computed.
Adding the $i\epsilon$ has no effect if these integrals are done
first, however when $i \epsilon$  is included it is possible to change 
the order of the time limit and the integration over the wave packets.  
I will reinsert the wave packets when I compute the cross section.  In deriving
(\ref{ap2.16}) the two strong limits in (\ref{ap2.11}) are replaced a
single weak limit.  Equation (\ref{ap2.11}) is interpreted as the
kernel of an integral operator.  $S$-matrix elements are obtained by
integrating the sharp eigenstates in Eq.~(\ref{ap2.16}) over
normalizable functions of the energy and other continuous variables.

To simply this expression I
define the residual interactions $V^\alpha$ and $V^\beta$ by:
\begin{equation}
V^\alpha := M - M_\alpha; \qquad V^\beta \:= M - M_\beta ,
\label{ap2.16}
\end{equation}
where
\begin{equation}
V^\alpha \vert \alpha \rangle = (M-{\sf W}_\alpha)
\vert \alpha \rangle; \qquad
V^\beta \vert
\beta \rangle =
(M-{\sf W}_\beta) \vert \beta \rangle.
\label{ap2.17}
\end{equation}
The resolvent operators of the mass operator and the
channel mass operator,
\begin{equation}
G(z) := {1 \over z-M} \qquad G_\alpha(z) := {1 \over z- M_\alpha},
\label{ap2.18}
\end{equation}
are related by the second resolvent relations \cite{Hi57}:
\begin{equation}
G(z) - G_\alpha(z) = G_\alpha(z)V^\alpha G(z) = G(z)V^\alpha G_\alpha(z).
\label{ap2.19}
\end{equation}
which when used in Eq.~(\ref{ap2.11}) gives
\begin{eqnarray}
\langle \beta \vert S \vert \alpha \rangle
&= &
\langle \beta \vert \alpha \rangle
\nonumber \\
&+ &
\lim_{\epsilon  \to 0^+} {1 \over 2} \langle \beta \vert
\left [ V^\beta \left (1 + G(\bar{\sf W}+ i \epsilon )V^\alpha  \right )
G_\alpha(\bar{\sf W}+i\epsilon) 
\right .
\nonumber \\
&+ &
\left .
G_\beta (\bar{\sf W}+ i \epsilon)\left
(1 + V^\beta G(\bar{\sf W}+i \epsilon)\right  )V^\alpha
\right ] \vert \alpha \rangle
\nonumber \\
&= &
\langle \beta \vert \alpha \rangle \left[ 1-\lim_{\epsilon  \to 0^+}
{{\sf W}_\beta - {\sf W}_\alpha \over {\sf W}_\beta -
{\sf W}_\alpha + 2i\epsilon } \right ]
\nonumber \\
 & &+
\lim_{\epsilon  \to 0^+}  \left [ {1 \over {\sf W}_\beta -
{\sf W}_\alpha +
2i\epsilon} + {1 \over {\sf W}_\alpha -{\sf W}_\beta + 2 i \epsilon } \right ]
\nonumber \\
&\times &
\langle \beta \vert \left( V^\alpha + V^\beta G(\bar{\sf W}+ i \epsilon )
V^\alpha  \right ) \vert \alpha\rangle
\nonumber \\
&= &
\langle \beta \vert \alpha \rangle \lim_{\epsilon  \to 0^+} \left
[ { 2i\epsilon \over {\sf W}_\beta - {\sf W}_\alpha + 2i\epsilon } \right ]
\nonumber \\
 &+&
\lim_{\epsilon  \to 0^+} \left  [ {-4i\epsilon \over ({\sf W}_\beta -
{\sf W}_\alpha)^2 + 4\epsilon^2} \right ] \langle \beta \vert
\left( V^\alpha + V^\beta G(\bar{\sf W}+ i \epsilon )V^\alpha
\right )\vert \alpha\rangle .
\label{ap2.20}
\end{eqnarray}
It is now possible to evaluate the limit as $\epsilon \to  0$.
It is important to remember that this is the kernel of an integral
operator.

The first term in square brackets is unity when the initial and final
mass eigenvalues are identical, and zero otherwise; however, the limit
in the bracket is a Kronecker delta and {\it not} a Dirac delta
function.  For $\alpha \not= \beta$, $\langle \beta ({\sf W}') \vert
\alpha ({\sf W} ) \rangle$ are Lebesgue measurable in ${\sf W}'$ for
fixed ${\sf W}$, so there is no contribution from the first term in
Eq. (\ref{ap2.20}).  For the case that ${\sf W}_\alpha = {\sf
W}_\beta$, we have $\langle \beta ({\sf W}') \vert \alpha ({\sf W})
\rangle\propto \delta ({\sf W}'-{\sf W} )$.  The matrix element
vanishes by orthogonality unless ${\sf W}_\beta = {\sf W}_\alpha$, but
then the coefficient is unity.  Thus, the first term in (\ref{ap2.20})
is $\langle \beta | \alpha \rangle$ if the initial and final channels
are the same, but zero otherwise.  The matrix elements also vanish by
orthogonality for two different channels governed by the same
asymptotic mass operator with the same invariant mass.  The first term
in (\ref{ap2.20})  therefore includes a {\it channel} delta function.

For the second term, the
quantity in square brackets becomes $-2 \pi i \delta ({\sf W}_\beta -
{\sf W}_\alpha)$, which leads to the relation
\begin{equation}
\langle \beta \vert S \vert \alpha \rangle = \langle \alpha \vert \beta \rangle
- 2 \pi i \delta ({\sf W}_\beta -
{\sf W}_\alpha) \langle
\beta \vert T^{\beta \alpha}({\sf W}_\alpha+i0^+) \vert \alpha \rangle,
\label{ap2.21}
\end{equation}
where
\begin{equation}
T^{\beta\alpha}(z) = V^\alpha + V^\beta G(z) V^\beta.
\label{ap2.22}
\end{equation}
and $\langle \alpha \vert \beta \rangle $ is zero
if the initial and final channels are different and is the overlap of
the initial and final states if the initial and final channels are the
same.  Equation (\ref{ap2.21}) is exactly eq. (\ref{ap2.11}).

The channel projection operators
$\Pi_\alpha$ are absorbed in the channel states, $\vert \alpha
\rangle$.  The translational invariance of the interaction
(\ref{ap2.16}) requires that
\begin{equation}
\langle \mathbf{P} , \cdots \vert
T^{\alpha \beta} (z) \vert \cdots ,  {\bf P}\,' \rangle =
\delta (\mathbf{P}- \mathbf{P}\,')
\langle \cdots \Vert T^{\alpha \beta} (z) \Vert \cdots \rangle .
\label{ap2.23}
\end{equation}

With our choice of irreducible basis the residual interactions and the
resolvent commute with the total linear momentum operator, and if the
sharp channel states $\vert \alpha \rangle$ and $\vert \beta \rangle$
are simultaneous eigenstates of the appropriate partition mass
operator and the linear momentum, then a three-momentum conserving
delta function can be factored out of the $T$-matrix element:
\begin{equation}
\langle \beta \vert T^{\beta\alpha}({\sf W}_\alpha+i0^+)
\vert \alpha \rangle =
\delta^3 ( \mathbf{P}_\beta - \mathbf{P}_\alpha )
\langle \beta \Vert  T^{\beta \alpha}({\sf W}_\alpha+i0^+)
\Vert  \alpha \rangle.
\label{ap2.24}
\end{equation}
When combined with the three-momentum conserving delta function
the invariant mass delta function can be replaced an energy
conserving delta function
\begin{equation}
\delta ({\sf W}_\beta - {\sf W}_\alpha) =
\left | {d{\sf W} \over d{\sf E}}  \right| \; \delta ({\sf E}_\beta - {\sf E}_\alpha)  \qquad 
\left| {d{\sf W} \over d{\sf E}}  \right| = 
{{\sf W} \over {\sf E}}
.
\label{ap2.25}
\end{equation}
The $S$-matrix elements
can be expressed in terms of the {\it reduced channel
transition operators} as follows:
\begin{equation}
\langle \beta \vert S \vert \alpha \rangle = \langle \alpha \vert \beta
\rangle
\delta_{\beta \alpha} - i (2 \pi) \delta^4 (P_\beta - P_\alpha)
{{\sf W}_{\alpha} \over {\sf E}_{\alpha}}
\langle \beta \Vert T^{\beta\alpha}({\sf W}_\alpha+i0^+
) \Vert  \alpha \rangle
\label{ap2.26}
\end{equation}
In this expression the $S$ operator is invariant while the single
particle asymptotic states have a non-covariant normalization.

\noindent{\bf Scattering cross sections}

The representation of the scattering matrix (\ref{ap2.26}) is used to 
calculate the cross section.  I derive the cross section
following standard
methods used by Brenig and Haag \cite{Br59}.  An initial state
consisting of a target $t$ in a state $\vert \varphi_t \rangle$
and beam $b$ in a state $\vert \varphi_b \rangle$ leads to
the asymptotic differential probability amplitude
for a $n$-particle final state in channel $\alpha$:
\begin{equation}
\langle \mathbf{p}_1, \cdots ,\mathbf{p}_n \vert \varphi \rangle :=
\int \langle \mathbf{p}_1, \cdots, \mathbf{p}_n
\vert S_{\alpha\beta} \vert
\mathbf{p}_b , \mathbf{p}_t \rangle d\mathbf{p}_b d\mathbf{p}_t
\langle \mathbf{p}_b \vert \varphi_b \rangle
\langle \mathbf{p}_t \vert \varphi_t \rangle .
\label{ap2.27}
\end{equation}
where the spin degrees of freedom are suppressed.
The differential probability for observing each final particle to be
within $d\mathbf{p}_i$ of $\mathbf{p}_i$ is
\begin{equation}
dP = \vert\langle \mathbf{p}_1, \cdots ,\mathbf{p}_n
\vert \varphi \rangle \vert^2 d \mathbf{p}_1
\cdots d \mathbf{p}_n .
\label{ap2.28}
\end{equation}
Inserting the expression (\ref{ap2.26}) for $S$ in terms of the wave
packets in (\ref{ap2.27}), assuming either different initial channels
or non-forward scattering, so there is no contribution from the
identity part of the S matrix, gives
\begin{eqnarray}
dP  &= &  d \mathbf{p}_1 \cdots d \mathbf{p}_n \int
(2\pi)^2 \langle \mathbf{p}_1, \cdots, \mathbf{p}_n
\Vert T^{\alpha \beta} \Vert
\mathbf{p}_b', \mathbf{p}_t' \rangle
\langle \mathbf{p}_1, \cdots, \mathbf{p}_n
\Vert T^{\alpha \beta} \Vert
\mathbf{p}_b'', \mathbf{p}_t'' \rangle^*   \nonumber \\
 & \times&
\delta
\left( \sum_i \mathbf{p}_i  - \mathbf{p}_b' -\mathbf{p}_t' \right)
 \; \delta \left( \sum_j \mathbf{p}_j  - \mathbf{p}_b'' -\mathbf{p}_t''
\right)   \nonumber \\
 & \times&
\delta ({\sf W}_\alpha - {\sf W}_{bt}' ) \;
\delta ({\sf W}_\alpha - {\sf W}_{bt}'' )
d \mathbf{p}_b' d \mathbf{p}_t' d \mathbf{p}_b'' d \mathbf{p}_t''
\nonumber \\
 & \times &
\langle \mathbf{p}_b' \vert \varphi_b \rangle  \langle \mathbf{p}_b''
\vert \varphi_b \rangle^*
\langle \mathbf{p}_t' \vert \varphi_t \rangle
\langle \mathbf{p}_t'' \vert \varphi_t \rangle^*.
\label{ap2.29}
\end{eqnarray}
The delta function for the conservation of linear momentum  means that
\begin{equation}
\delta ({\sf W}_\alpha - {\sf W}_{bt}' )=
\delta ({\sf E}_\alpha - {\sf E}_{bt'}) \left| {dE \over dM}\right| =
\delta ({\sf E}_\alpha - {\sf E}_{bt}')
\left| {{\sf W}_\alpha  \over {\sf E}_\alpha}\right|
\label{ap2.30}
\end{equation}
where ${\sf E}_\alpha=\sqrt{{\sf W}_{\alpha}^2 + \mathbf{P}^2}$.
With this replacement the delta functions in (\ref{ap2.30}) can
be replaced by products of four momentum conserving delta functions:
\[
\delta^4 ( \sum_i p_i - p_b' -p_t') \; \delta^4 ( \sum_j p_j - p_b'' -p_t'') =
\]
\begin{equation}
\delta^4 ( \sum_i p_i - p_b' -p_t') \; \delta^4 (p_b' + p_t' - p_b'' -p_t'').
\label{ap2.31}
\end{equation}
{\color{red} If the initial wave packets are sharply peaked about the
target and beam momenta and the transition operator varies slowly on
the support of these wave packets, then the transition operators
can be factored out of the integral, replacing the momenta with the
mean target and beam momenta, $\bar{\mathbf{p}}_b,
\bar{\mathbf{p}}_t$.  This approximation must be valid for the cross
section to be independent of the shape of the wave packets.}  The
result, after expressing the second four momentum conserving delta
function in (\ref{ap2.31}) by a Fourier integral representation
\begin{equation}
{1 \over {2\pi}^4} \int e^{i x \cdot (p_b' + p_t'  -p_b'' -p_t'')}
d^4x
\label{ap2.32}
\end{equation}
is
\begin{eqnarray}
dP  &= & (2\pi)^{4} d \mathbf{p}_1 \cdots  d \mathbf{p}_n \,
\int \left|  \langle \mathbf{p}_1, \cdots,
\mathbf{p}_n \Vert T^{\alpha \beta} \Vert
\bar{\mathbf{p}}_b,  \bar{\mathbf{p}}_t  \rangle \right| ^2
\vert {{\sf W}_\alpha  \over {\sf E}_\alpha} \vert^2
\nonumber \\
&\times &
\vert \langle \mathbf{x}, t \vert \varphi_b \rangle\vert^2
\vert \langle \mathbf{x}, t \vert \varphi_t \rangle\vert^2
d\mathbf{x}  \; dt
\nonumber \\
&\times &
\delta \left( \sum_i \mathbf{p}_i  -
\bar{\mathbf{p}}_b -\bar{\mathbf{p}}_t \right) \;
\delta \left( \sum_i {\sf E}_{k_i}  - \bar{\sf E}_b -\bar{\sf E}_t \right).
\label{ap2.33}
\end{eqnarray}
This is the differential probability for a single scattering event.
The space-time integral picks up a contribution whenever the beam and
target are in the same place at the same time.

In a real experiment there is a statistical ensemble of $N_t$ target
particles with number density
\begin{equation}
\rho_t (\mathbf{x},t) =
N_t \sum_l p_{lt}  \vert \langle \mathbf{x}, t \vert \varphi_{lt}
\rangle\vert^2
\label{ap2.34}
\end{equation}
where $N_t$ is the total number of target particles and $p_{lt}$ are
probabilities in the target density matrix.
A normal beam current density with a flux of
$N_b$
beam particles per unit time per unit cross sectional area
can be expressed as
\begin{equation}
j_b (\mathbf{x},t) = v_{bt} \rho_b (\mathbf{x},t) =
v_{bt}
N_b \sum_m p'_{mb}  \vert \langle \mathbf{x}, t \vert \varphi_{mt}
\rangle\vert^2
\label{ap2.35}
\end{equation}
where $p'_{mb}$ are probabilities in the beam density matrix and
$v_{bt}$ is the relative speed of the beam with respect to the
target.

Using this in the expression for the scattering probability,
assuming that there is no more than one scattering event per
incident particle,
gives the number of particles scattered per unit time per unit
volume as
\begin{eqnarray}
{dN \over d^4x} &=&  N_b N_t {dP \over d\mathbf{x} dt}  =
{(2\pi)^{4}\over v_{bt} }
\left |  \langle \mathbf{p}_1, \cdots, \mathbf{p}_n
\Vert T^{\alpha \beta} \Vert
\bar{\mathbf{p}_1}_b,  \bar{\mathbf{p}_1}_t  \rangle \right| ^2
{{\sf W}_\alpha^2  \over {\sf E}^2_\alpha}
\nonumber \\
 &  \times &
\delta^4 (p_1 + \cdots + p_n - \bar{p}_b -\bar{p}_t) \;
d\mathbf{p}_1 \cdots d\mathbf{p}_n \;
\rho_t (\mathbf{x},t) j_b (\mathbf{x},t) .
\label{ap2.36}
\end{eqnarray}
This quantity is proportional to the beam current times the
target density.  The proportionality factor is the 
differential cross section, $d\sigma$.  Comparing 
\beq
{dN \over d^4x} = d\sigma \rho_t (\mathbf{x},t) j_b (\mathbf{x},t) 
\label{ap2.37}
\eeq
with (\ref{ap2.36}) 
gives the following expression for the differential cross section:
\[
{d\sigma} =
{(2\pi)^{4}\over v_{bt}}
\left|  \langle \mathbf{p}_1, \cdots, \mathbf{p}_n
\Vert T^{\alpha \beta} \Vert
\bar{\mathbf{p}}_b,  \bar{\mathbf{p}}_t  \rangle \right| ^2
\times
\]
\begin{equation}
{{\sf W}_\alpha^2  \over {\sf E}^2_\alpha}
\;
\delta^4 \left( \sum_{i=1}^n p_i - \bar{p}_b -\bar{p}_t \right) \;
d\mathbf{p}_1 \cdots d\mathbf{p}_n  .
\label{ap2.38}
\end{equation}
For identical particles this must be multiplied by a statistical 
factor ${1 \over s}$, where $s$ is the number of permutations of the 
identical particles in the final state (i.e. $s=n_1! n_2! \cdots$ if there 
$n_1$ identical particle of type 1, $n_2$ identical particle of 
type 2, etc. in the final state.)

For polarized beams or targets it is useful to 
introduce matrices  $S_{ia}$ and $S_{fa}$ in the initial 
and final spin variables with the normalization
\beq
\mbox{Tr} (S_{ia} S_{ib}) = \delta_{ab} \qquad   
\mbox{Tr} (S_{fa} S_{fb}) = \delta_{ab}  
\eeq
Spin operators are linear combinations of these operators 
with constant coefficients $s_{ia}$ and $s_{fa}$ 
\beq
S_f = \sum_a s_{fa}S_{fa}  \qquad 
S_i = \sum_a s_{ia}S_{ia}  
\eeq
the resulting observable is 
\beq
\langle O \rangle = 
{\mbox{Tr} ( T^{\dagger}  S_f T S_i )  \over 
\mbox{Tr} ( T^{\dagger}   T  )}  
\eeq
where the traces are over the spins.  All of the spin independent 
factors cancel in the ratio.  In general the choice of spin-basis matrices
$S$ depend on the particle content of the initial and or final states.

Except for the factor ${\sf W}_\alpha^2 / {\sf E}^2_\alpha$ Eq.~(\ref{ap2.38}) 
is identical to the corresponding non-relativistic
expression.  The additional factor of ${\sf W}_\alpha^2 / {\sf
E}^2_\alpha$ arises because we have chosen to calculate the
transition operator using the mass operator instead of the
Hamiltonian.
The only difference in these formulas from standard formulas is that
the transition operator is constructed from the difference of the mass
operators with and without interactions and the appearance of the
additional factor of ${\sf W}_\alpha^2 / {\sf E}^2_\alpha$ which
corrects for the modified transition operator.  This factor becomes 1
when $\mathbf{P}=0$.

The expression of Eq.~(\ref{ap2.28}) can be expressed in a manifestly
invariant form.   The relation to the standard expression of the invariant
cross section using conventions of the particle data book \cite{pdg}
is derived below.

\section{Invariance of $S$ and relation to $T$}
\label{appendixa}

The expression (\ref{ap2.38}) for the differential cross section
can be rewritten in a manifestly invariant form.  We write it as a
product of an invariant phase space factor, an invariant factor
that includes the relative speed, and an invariant scattering amplitude.

To establish the invariance of the invariant scattering
amplitude note that the scattering operator $S$ is Poincar\'e invariant:
\begin{eqnarray}
U_f (\Lambda ,a) S &=&
U_f (\Lambda ,a) \Omega_+^{\dagger} (H,H_0)
\Omega_- (H,H_0) 
\nonumber \\
&=&
=
\Omega_+^{\dagger} (H,H_0) U (\Lambda ,a)
\Omega_- (H,H_0) 
\nonumber \\
&=&\Omega_+^{\dagger} (H,H_0)
\Omega_- (H,H_0) U_f (\Lambda ,a) =
S U_f (\Lambda ,a).
\label{ap2.39}
\end{eqnarray}
where $U_f$ is the product of the cluster irreducible representations of the
Poincar\'e group that act on the channel states. 

The proof of the Poincar\'e invariance of the $S$ operator above is a
consequence of the intertwining relations for the wave operators
\begin{equation}
U (\Lambda ,a)
\Omega_\pm (H,H_0) =
\Omega_\pm (H,H_0) U_0 (\Lambda ,a)
\label{ap2.40}
\end{equation}
To show the intertwining property of the wave operators first note
that the invariance principle gives the identity
\begin{equation}
\Omega_\pm (H,H_0) = \Omega_\pm (M,M_0) .
\label{ap2.41}
\end{equation}
The mass operator intertwines by the standard intertwining properties
of wave operators.  For our choice of irreducible basis the
intertwining of the full Poincar\'e group follows because all of
the generators can be expressed as functions of the mass
operator and a common set of kinematic operators, $\{ \mathbf{P}, j_z
, j_x, j^2, -i\mbox{\boldmath$\nabla$}_P\}$, that commute with the
wave operators.

The covariance of the $S$ matrix elements follows from the Poincar\'e
invariance of the $S$ operator if the matrix elements of $S$ are
computed in a basis with a covariant normalization.

To extract the standard expression for the invariant amplitude the
single particle states are replaced by states with the covariant
normalization used in the particle data book \cite{pdg}:

\begin{equation}
\vert \mathbf{p}, \mu \rangle \longrightarrow  \vert p, \mu \rangle_{cov} =
\vert \mathbf{p},\mu \rangle \sqrt{2 E_{k_m}}(2 \pi)^{3/2} .
\label{ap2.42}
\end{equation}
The resulting expression
\begin{equation}
- i (2 \pi) \delta^4 (P_\beta - P_\alpha)
{{\sf W}_{\alpha} \over {\sf E}_{\alpha}}{}
_{cov}\langle \beta \Vert T^{\beta\alpha}({\sf W}_\alpha+i0^+)
\Vert \alpha \rangle_{cov}
\label{ap2.43}
\end{equation}
is invariant (up to spin transformation properties).  Since the four
dimensional delta function is invariant, the factor multiplying the
delta function is also invariant (up to spin transformation properties).
This means that
\begin{equation}
_{cov}\langle \alpha \Vert M^{\alpha \beta} \Vert \beta \rangle_{cov} :=
{1 \over (2 \pi)^3}
{{\sf W}_{\alpha} \over {\sf E}_{\alpha}}{}
_{cov}\langle \beta \Vert T^{\beta\alpha}({\sf W}_\alpha+i0^+) \Vert
\alpha \rangle_{cov}
\label{ap2.44}
\end{equation}
is a Lorentz covariant amplitude.  The factor of $1/(2 \pi)^3$ is chosen
to agree with the normalization convention used in the particle
data book \cite{pdg}.

The differential cross
section becomes
\begin{eqnarray}
{d \sigma } & = &
{(2\pi)^4 \over
4  E_{m_t} (\mathbf{p}_t)
E_{m_b} (\mathbf{p}_b) v_{bt}} \;
\left|_{cov}\langle p_1, \cdots , p_n, \Vert M^{\alpha \beta} \Vert
\bar{p}_b,  \bar{p}_t  \rangle_{cov} \right| ^2
\nonumber \\
 &\times &
\delta^4  \left( \sum_i p_i  - \bar{p}_b -\bar{p}_t \right) \;
{d\mathbf{p}_1 \over  2 E_{m_1} (2 \pi)^3 }
\cdots
{d\mathbf{p}_n \over  2 E_{m_n} (2 \pi)^3 }.
\label{ap2.45}
\end{eqnarray}
The identity
\begin{equation}
v_{bt}  ={\sqrt{ p_t\cdot p_b)^2 -m_b^2 m_t^2} \over
E_{m_b} E_{m_t} }
\label{ap2.46}
\end{equation}
can be used to get an invariant expression for the relative speed between the
projectile and target and
\[
d\Phi_n ({p}_b+{p}_t; \mathbf{p}_1, \cdots , \mathbf{p}_n) =
\delta^4 \left( \sum_i p_i - \bar{p}_b -\bar{p}_t \right)
\]
\begin{equation}
\times \;
{d\mathbf{p}_1 \over 2 E_{m_1} (2 \pi)^3 } \cdots
{d\mathbf{p}_n \over 2  E_{m_n} (2 \pi)^3 }
\label{ap2.47}
\end{equation}
is the standard Lorentz invariant phase space factor.  Inserting these
covariant expressions in the definition of the differential cross
section gives the standard formula for the invariant cross section
\[
d \sigma =
{(2 \pi)^4 \over
4 \sqrt{ (p_t\cdot p_b)^2 -m_b^2 m_t^2}}
\left|_{cov}\langle p_1, \cdots , p_n, \Vert M^{\alpha \beta} \Vert
\bar{p}_b,  \bar{p}_t  \rangle_{cov} \right| ^2 
\]
\begin{equation}
\times
d\Phi_n ({p}_b+{p}_t; \mathbf{p}_1, \cdots , \mathbf{p}_n).
\label{ap2.48}
\end{equation}
Because of the unitarity of the Wigner rotations and the covariance of
\beq
\left|_{cov}\langle p_1, \cdots , p_n, \Vert M^{\alpha \beta} \Vert
\bar{p}_b, \bar{p}_t \rangle_{cov} \right| ^2
\label{ap2.49}
\eeq
this becomes an
invariant if the initial spins are averaged and the final spins are
summed.  In our model with spinless nucleon the total cross section is
invariant.

This manifestly invariant formula for the cross section is identical
to (\ref{ap2.38}); in this form the invariant cross section can
be evaluated in any frame. The index $t$ refers to the target, which is
in our case the deuteron.

This work was performed under the auspices of the U.~S.  Department of
Energy,  Office of Nuclear Physics, under contract
No. DE-FG02-86ER40286.  The author would like to express his gratitude
to Franz Gross and the TJNAF theory group for the invitation to
present these lectures.


\begin{thebibliography}{00}

\bibitem{clay} http://www.claymath.org/millennium/

\bibitem{St65} R. F. Streater and A. S.  Wightman, {\it PCT, Spin And
Statistics, And All That}, Benjamin/Cummings, Reading, Mass. (1964),
Theorem 4.1.

\bibitem{fg} Franz L. Gross,
``Relativistic Quantum Mechanics and Field Theory'' 
(1999) - Science.

\bibitem{schroer} Bert Schroer, arXiv:hep-th/0405105v4, 2004;
arXiv:0711.4600v2 [hep-th], 2007

\bibitem{baum}  H. Baumg\"artel and M. Wollenberg,
{\it Mathematical Scattering Theory}, Birkhauser, 1983.

\bibitem{Wi39} E. P. Wigner, {\it Ann. Math.} {\bf 40},149(1939).

\bibitem{Haag} R. Haag, ``Local Quantum Physics'', Springer, 1982.

\bibitem{Os} K. Osterwalter and R. S. Schrader, {\it Comm. Math. Phys.} 
{\bf 42},281(1975)

\bibitem{Ba54} V. Bargmann, {\it Ann. Math.} {\bf 59},1 (1954).

\bibitem{Ne49} T.D.Newton and E.P.Wigner, {\it Rev. Mod. Phys.} {\bf
21},400(1949).

\bibitem{Wi60} A. S. Wightman in
{\it Relations de Dispersion et Particules \'El\'ementaries}, ed. C. De
Witt and R. Omn\'es,  Hermann, Paris (1960). 

\bibitem{Lu42} L.Lubanski, {\it Physica (Utrecht)} {\bf 9},310(1942).

\bibitem{Ba53} B. Bakamjian and L. H. Thomas, {\it Phys. Rev.} {\bf
92},1300 (1953).

\bibitem{Co65} F. Coester, {\it Helv. Phys. Acta}, {\bf 38},7(1965).

\bibitem{So77} S. N. Sokolov, {\it Sov. Phys. Dokl.}
{\bf 233},575(1977). 

\bibitem{Co82} F. Coester and W. N. Polyzou, {\it Phys. Rev. D} {\bf
26},1348(1982).
 
\bibitem{dirac} P. A. M. Dirac, {\it Rev. Mod. Phys}, {\bf 21},392(1949).

\bibitem{wpwk} W. Polyzou and W. Klink, {\it Ann. Phys.} {\bf 185},369(1988). 

\bibitem{Wi75} K. G. Wilson, Erice Lectures, 1975, edited by A. Zichichi, 
Plenum Press, New York.

\bibitem{Seiler} E. Seiler, ``Gauge theories as a problem in 
constructive quantum field theory and statistical mechanics'', Springer 1982. 

\bibitem{dashen} R. F. Dashen, J. B. Healy, I. J. Muzinich, 
Ann. Physics, 102,1,(1976).

\bibitem{Bo59} N. N. Bogoliubov and D. V.  Shirkov, ``Introduction to
  the Theory of Quantized Fields'' (Interscience Monographs in Physics
  and Astronomy) 1959.
	
\bibitem{Me74} H. J. Melosh, {\it Phys. Rev. D} {\bf 9},1095(1974).

\bibitem{keipo} B.D. Keister, W. N. Polyzou, Phys. Rev. C {\bf 73}, 014005
(2006).

\bibitem{Ch76} C. Chandler and A. Gibson, {\it Indiana J. Math.} {\bf
25},443(1976).

\bibitem{Hi57} E. Hille and R. S. Phillips, {\it Functional Analysis and
Semigroups}, American Mathematical Society Colloquium Publications, Vol
31, Providence, Rhode Island (1957), sec 4.8

\bibitem{Br59} W. Brening and R. Haag, {\it  Fort. der. Physik}, {\bf
7},183(1959).

{\it Mathematical Scattering Theory}, Birkhauser, 1983.

\bibitem{pdg} Particle Data Group - 
W.-M. Yao et al., Journal of Physics, G 33, 1 (2006).

\end{thebibliography}
\end{document}